\documentclass{emulateapj}
\usepackage{apjfonts}

\usepackage{ifpdf}
\ifpdf
\else
\fi

\newcommand{\lya}{Ly$\alpha$}
\newcommand{\ha}{H$\alpha$}
\newcommand{\hb}{H$\beta$}
\newcommand{\oiii}{[O III]}

\newcommand{\msun}{\mbox{M$_{\odot}$}}
\newcommand{\fluxcgs}{erg s$^{-1}$ cm$^{-2}$}
\def\arcsec{^{\prime\prime}}

\def\z{z}
\def\nh{$N_{\mathrm{HI}}$}
\def\vexp{$V_{\mathrm{exp}}$}
\def\chisq{$\chi^{2}$}
\def\ewlya{EW$_{\mathrm{Ly}\alpha}$}
\def\hii{H\,{\sc ii}}

\def\apj{ApJ}
\def\apjs{ApJS}
\def\asp{ASP Conf. Ser.}
\def\mnras{MNRAS}
\def\araa{ARA\&A}
\def\aj{AJ}
\def\aap{A\&A}
\def\nat{Nature}
\def\pasp{PASP}
\def\spie{Proc. SPIE}

\begin{document}
\slugcomment{(Accepted for publication in \textit{The Astrophysical Journal}; August 5, 2013)}

\shorttitle{SPECTRALLY RESOLVED \lya\ EMISSION IN $\z \sim 2.4$ FIELD GALAXIES}
\shortauthors{T. S. CHONIS ET AL.}

\title{The Spectrally Resolved \lya\ Emission of Three \lya-Selected Field Galaxies at $\z \sim 2.4$ from the HETDEX Pilot Survey\altaffilmark{*}}

\author{Taylor S. Chonis\altaffilmark{1}, Guillermo A. Blanc\altaffilmark{2}, Gary J. Hill\altaffilmark{3}, Joshua J. Adams\altaffilmark{2}, Steven L. Finkelstein\altaffilmark{1}, Karl Gebhardt\altaffilmark{1}, Juna A. Kollmeier\altaffilmark{2}, Robin Ciardullo\altaffilmark{4,5}, Niv Drory\altaffilmark{3}, Caryl Gronwall\altaffilmark{4,5}, Alex Hagen\altaffilmark{4,5}, Roderik A. Overzier\altaffilmark{1,6}, Mimi Song\altaffilmark{1}, and Gregory R. Zeimann\altaffilmark{4,5}}

\altaffiltext{*} {This paper includes data taken at The McDonald Observatory of The University of Texas at Austin.}
\altaffiltext{1} {Dept. of Astronomy, University of Texas at Austin, 2515 Speedway, Stop C1400, Austin, TX 78712, USA: tschonis@astro.as.utexas.edu}
\altaffiltext{2} {Observatories of the Carnegie Institution of Washington, 813 Santa Barbara Street, Pasadena, CA 91101, USA}
\altaffiltext{3} {McDonald Observatory, University of Texas at Austin, 2515 Speedway, Stop C1402, Austin, TX 78712, USA}
\altaffiltext{4} {Dept. of Astronomy \& Astrophysics, The Pennsylvania State University, 525 Davey Lab, University Park, PA 16802, USA}
\altaffiltext{5} {Institute for Gravitation and the Cosmos, The Pennsylvania State University, 104 Davey Lab \#258, University Park, PA 16802, USA}
\altaffiltext{6} {Observat\'{o}rio Nacional, Rua Jos\'{e} Cristino, 77. CEP 20921-400, S\~{a}o Crist\'{o}v\~{a}o, Rio de Janeiro-RJ, Brazil}

\begin{abstract}
We present new results on the spectrally resolved \lya\ emission of three \lya\ emitting field galaxies at $\z\sim2.4$ with high \lya\ equivalent width ($>100$ \AA) and \lya\ luminosity ($\sim10^{43}$ erg s$^{-1}$). At 120 km s$^{-1}$ (FWHM) spectral resolution, the prominent double-peaked \lya\ profile straddles the systemic velocity, where the velocity zero-point is determined from spectroscopy of the galaxies' rest-frame optical nebular emission lines. The average velocity offset from systemic of the stronger redshifted emission component for our sample is 176 km s$^{-1}$ while the average total separation between the redshifted and main blueshifted emission components is 380 km s$^{-1}$. These measurements are a factor of $\sim2$ smaller than for UV continuum-selected galaxies that show \lya\ in emission with lower \lya\ equivalent width. We compare our \lya\ spectra to the predicted line profiles of a spherical ``expanding shell'' \lya\ radiative transfer grid that models large-scale galaxy outflows. Specifically blueward of the systemic velocity where two galaxies show a weak, highly blueshifted (by $\sim1000$ km s$^{-1}$) tertiary emission peak, the model line profiles are a relatively poor representation of the observed spectra. Since the neutral gas column density has a dominant influence over the shape of the \lya\ line profile, we caution against equating the observed \lya\ velocity offset with a physical outflow velocity, especially at lower spectral resolution where the unresolved \lya\ velocity offset is a convoluted function of several degenerate parameters. Referring to rest-frame ultraviolet and optical \textit{Hubble Space Telescope} imaging, we find that galaxy-galaxy interactions may play an important role in inducing a starburst that results in copious \lya\ emission, as well as perturbing the gas distribution and velocity field which have strong influence over the \lya\ emission line profile.\\ 
\end{abstract}

\section{INTRODUCTION}\label{sec:Introduction}
As the strongest spectral feature of the most abundant element in the universe, the \lya\ transition of hydrogen ($n = 2 \rightarrow 1$; 1216 \AA) has become one of the most widely utilized tools for detecting galaxies in the early universe (e.g., \citealp{hu1996,cowie1998,rhoads2000,steidel2000,vanbreukelen2005,gronwall2007,finkelstein2009,rauch2011,adams2011}). \lya\ is versatile in that it can probe the extremes of the high-redshift galaxy census, such as faint, low-mass populations (e.g., \citealp{rauch2008}) and the high energy systems at the top-end of the mass function (e.g., \citealp{willott2011}). 

\lya\ photons are produced as a result of excitation or ionization of neutral hydrogen and can arise in a variety of astrophysical scenarios relating to high-redshift galaxies. These include cooling radiation as a result of gravitational collapse (e.g., \citealp{faucher-giguere2010}) or extremely rapid gas motions (i.e., shocks; \citealp{birnboim2003}), photo-ionization by stars (e.g., \citealp{partridge1967}) or an active galactic nucleus (AGN; e.g., \citealp{haiman2001}), or fluorescence of neutral gas in the intergalactic medium (IGM; e.g., \citealp{kollmeier2010}). However, the interpretation of the emergent \lya\ spectra is non-trivial due to the resonant nature of the transition in neutral hydrogen, in which the line is optically thick for even small neutral fractions \citep{gunn1965}. This property makes the escape of \lya\ photons from a galaxy a complex radiative process that is highly dependent on the properties of the interstellar medium (ISM) and surrounding circumgalactic medium, such as gas kinematics, covering fraction, geometry, and dust content. The result of \lya\ radiative transfer through the ISM is a significant modulation of the intrinsic \lya\ emission line profile, both spatially and spectrally. The emergent emission line profile can be further modified by transfer through the surrounding IGM, which can absorb a significant amount of radiation in the vicinity and blueward of \lya\ (e.g., \citealp{laursen2011}). Because of the amount of physics that can be encoded within the emergent \lya\ line profile, a great deal of theoretical work has gone into understanding \lya\ radiative transfer to aid in the interpretation of observed \lya\ spectra of galaxies. In particular, the use of Monte Carlo numerical techniques has become standard practice (e.g., \citealp{lee1974,ahn2000,zheng2002,tasitsiomi2006,hansen2006,dijkstra2006a,verhamme2006}), and has provided useful predictions of the emergent \lya\ line profile for comparison with observations for a variety of different physical scenarios. The implementation of such radiative transfer codes are currently being advanced by post-processing more realistic models of galaxies drawn from hydrodynamic and cosmological simulations (e.g., \citealp{zheng2010,kollmeier2010,barnes2011,verhamme2012,yajima2012}).

Non-active galaxies that are selected by virtue of their \lya\ emission with equivalent width \ewlya\ $> 20$ \AA\ (i.e., \lya\ emitters; LAEs) are typically low-mass ($\lesssim10^{9.5}$ \msun) with young stellar populations ($\lesssim100$ Myr) and are forming stars at rates of $\lesssim10$ \msun\ yr$^{-1}$ (e.g., \citealp{venemans2005,gawiser2006,finkelstein2007,ono2010,acquaviva2011}). In addition, some LAEs appear to have non-negligible dust content \citep{finkelstein2009,pentericci2009}, low gas-phase metallicity \citep{finkelstein2011,nakajima2013}, and compact, irregular spatial morphologies (e.g., \citealp{venemans2005,gronwall2011,bond2012}). At least some of these properties appear to not evolve significantly with redshift (e.g., \citealp{ouchi2008,blanc2011,mallery2012}). A significant fraction of ultraviolet (UV) continuum-selected galaxies (i.e., Lyman break galaxies; LBGs) also show \lya\ in emission (e.g., \citealp{shapley2003,kornei2010,stark2010,kulas2012,ono2012,berry2012}). LBGs typically have lower \ewlya\ (some show \lya\ in absorption), more evolved stellar populations, and are more massive than the typical \lya-selected galaxy (e.g., \citealp{gawiser2006,pentericci2007,yuma2010}).

In a star-forming galaxy, \lya\ photons are produced from recombinations in hydrogen gas that was ionized by the UV radiation of massive, main sequence stars in \hii\ regions. Unlike UV continuum or photons from other optically thin transitions that hail from the same \hii\ region (e.g., \ha), \lya\ photons become resonantly trapped in the first parcels of neutral hydrogen they encounter. As compared to the optically thin photons, the path length of \lya\ to escape the galaxy is increased due to the scattering, making it especially sensitive to dust absorption. The path length of \lya\ photons through neutral hydrogen can be further altered by local velocity fields due to thermal motion, turbulence, and bulk motions (e.g., due to supernovae or stellar driven winds, or by galaxy-galaxy interactions), thereby shifting \lya\ photons in and out of resonance and affecting the emergent line profile (e.g., \citealp{verhamme2006}). This potentially makes the emergent \lya\ spectrum a very powerful probe of gas kinematics (e.g., \citealp{verhamme2008,yang2011,kulas2012}), especially when the spectrum can be spatially resolved (e.g., \citealp{rauch2011}).

Evidence for large-scale outflows is wide-spread in high-redshift star forming galaxies (e.g., \citealp{shapley2003,martin2005,berry2012}). In LBGs, for example, the galaxy's redshift differs when measured from the interstellar (IS) absorption lines versus the \lya\ emission line (by $\gtrsim600$ km s$^{-1}$), which implies that one or both features are not at rest with respect to the stars in the galaxy \citep{shapley2003}. Systemic redshifts and gravitationally induced motion are expected to be well represented by nebular emission lines such as \ha, which is optically thin and whose strength depends upon the UV radiation field in the same vicinity as the origin of the \lya\ photons \citep{erb2006}. In LBGs, the IS absorption lines are blueshifted with respect to the \ha-based systemic redshift (by $\sim160$ km s$^{-1}$) while the peak of the \lya\ emission line is typically redshifted (by $\sim450$ km s$^{-1}$; \citealp{steidel2010}). Although the interpretation of the redshifted \lya\ emission is non-trivial due to the transition's resonant scattering nature, the blueshifted IS absorption is a strong indicator of a galactic outflow and is caused by the absorption of stellar light in swept-up material that is approaching along the observer's line of sight.

	\begin{deluxetable*}{cccccl}
	\tabletypesize{\scriptsize}
	\tablecaption{OBSERVED PROPERTIES OF TARGETED HETDEX PILOT SURVEY GALAXIES\label{table:GalData}}
	\tablehead{ & \colhead{(Units)} & \colhead{HPS194} & \colhead{HPS256} & \colhead{HPS251} & Source\tablenotemark{$a$}\\}
	\startdata
	R.A. & (J2000) & 10:00:14.18 & 10:00:28.33 & 10:00:27.23 & A11\tablenotemark{$b$}\\
	Decl. & (J2000) & +02:14:26.11 & +02:17:58.44 & +02:17:31.50 & \\[2.5ex]
	$z_{\mathrm{Ly}\alpha}$ & ( - ) & $2.2896\pm0.0004$ & $2.4923\pm0.0004$ & $2.2865\pm0.0004$ & A11\tablenotemark{$c$}\\[0.5ex]
	$z_{\mathrm{sys}}$ & ( - ) & $2.28628\pm0.00002$ & $2.49024\pm0.00004$ & $2.28490\pm0.00005$ & F11\tablenotemark{$d$}, S13\tablenotemark{$d$}\\[0.5ex]
	$\Delta v_{\mathrm{Ly}\alpha}$ & (km s$^{-1}$) & $303\pm28$ & $177^{+52}_{-68}$ & $146^{+116}_{-156}$ & A11\tablenotemark{$c$},F11\tablenotemark{$d$}, S13\tablenotemark{$d$}\\[0.5ex]
	$F_{\mathrm{Ly}\alpha}$ & ($10^{-17}$ ergs s$^{-1}$ cm$^{-2}$) & $61.0^{+4.9}_{-4.3}$ & $31.4^{+9.3}_{-6.5}$ & $45.0^{+13.7}_{-11.6}$ & A11\\[0.5ex]
	$L_{\mathrm{Ly}\alpha}$ & ($10^{42}$ ergs s$^{-1}$) & $25.1\pm1.9$ & $16.4\pm4.1$ & $18.5\pm5.2$ & \\[0.5ex]
	EW$_{\mathrm{Ly}\alpha}$ & (\AA) & $114\pm13$ & $206\pm65$ & $140\pm43$ & B11\\[0.5ex]
	$\sigma_{\mathrm{H}\alpha}$ & (km s$^{-1}$) &  $61\pm9$ & $72\pm9$ & $44\pm6$ & F11, S13\\[0.5ex]
	$V$ & (mag) & $24.07\pm0.05$ & $25.07\pm0.09$ & $24.70\pm0.07$ & COSMOS\tablenotemark{$e$}\\[0.5ex]
	$E(B-V)$ & (mag) & $0.09\pm0.06$ & $0.10\pm0.09$ & $0.07\pm0.08$ & B11\\[0.5ex]
	SFR(H$\alpha$) & (\msun\ yr$^{-1}$) & $>29.3$ & $>35.4$ & $>9.9$ & F11, S13\tablenotemark{$f$}\\[0.5ex]
	$12 + \log{(\mathrm{O/H})}$ & ( - ) & $<7.87$ & $<8.12$ & $<8.00$ & F11, S13\tablenotemark{$g$}\\[0.5ex]
	Stellar Mass & ($10^{8}$ \msun) & $155^{+31}_{-43}$ & $1.9^{+0}_{-0.1}$ & $11^{+2}_{-1}$ & F11, S13\\
	\enddata
	\tablenotetext{$a$}{A11 = \citet{adams2011}, F11 = \citet{finkelstein2011}, S13 = Song et al. (2013, in preparation), B11 = \citet{blanc2011}}
	\tablenotetext{$b$}{The equatorial coordinates correspond to the optical counterpart to the \lya\ emission.}
	\tablenotetext{$c$}{Given for the HPS $R\approx750$ \lya\ spectra after the correction for $V_{LSR}$ and for the $n_{atm}$ error (see $\S$\ref{subsec:HPSoffset}). The \lya\ velocity offset is calculated from these data as: $\Delta v_{\mathrm{Ly}\alpha} = c\:(z_{\mathrm{Ly}\alpha} - z_{\mathrm{sys}})\: /\: (z_{\mathrm{sys}} + 1)$. The quoted uncertainties are statistical. Systematic errors associated with the wavelength calibration are discussed in $\S$\ref{subsec:HPSoffset}.}
	\tablenotetext{$d$}{Given for the NIRSPEC data after recalculation using the proper vacuum wavelengths for \oiii\ and \ha\ (see $\S$\ref{subsubsec:NIRSPEC}). For HSP251, $z_{\mathrm{sys}}$ is determined from \ha\ only. The quoted uncertainties are statistical. Systematic errors associated with the wavelength calibration are discussed in $\S$\ref{subsubsec:NIRSPEC}.}
	\tablenotetext{$e$}{COSMOS Intermediate and Broad Band Photometry Catalog, April 2009 Release. The magnitudes have been corrected to total using the supplied band-independent aperture corrections. Also, see \citet{ilbert2009}.}
	\tablenotetext{$f$}{Given from the measured \ha\ line fluxes \textit{without a dust correction} and assuming the star formation scaling relation of \citet{kennicutt1998}.}
	\tablenotetext{$g$}{The oxygen abundances are given as 1$\sigma$ upper limits.}
	\end{deluxetable*}

Recently, observations of rest-frame optical nebular emission lines (e.g., \hb, \oiii, or \ha, depending on the redshift of the galaxy) for $2 \lesssim z \lesssim 3$ LAEs observed in the near-infrared (NIR) have become more common. These data provide not only standard rest-frame optical emission line diagnostics for these high-redshift galaxies (e.g., \citealp{finkelstein2011,song2013}; Song et al. 2013, in preparation), but also a measure of their systemic redshifts \citep{mclinden2011,hashimoto2012,guaita2013}. Thus, \lya\ velocity offsets are now be measured for LAEs much in the same way as for LBGs (e.g., \citealp{steidel2010,kulas2012}), which provides an additional constraint when comparing \lya\ radiative transfer models to the observed \lya\ spectra of LAEs. However, velocity offsets of \lya\ alone cannot necessarily constrain the gas kinematics and may not correlate directly with bulk gas velocities. Useful kinematic information can be encoded within the offset \lya\ line profile in the form of asymmetries and other finer-scale spectral features, such as multiple emission peaks. This information can potentially be lost when observed at low spectral resolution, as has been used in multiple studies that probe \lya\ velocity offsets (e.g., \citealp{finkelstein2011,mclinden2011,kulas2012,guaita2013}; Song et al. 2013, in preparation). Multiple peaks are a natural consequence of \lya\ resonant line transfer (e.g., \citealp{neufeld1990}), and the frequency of multiple-peaked \lya\ spectra in LAE and LBG samples has recently been studied by \citet{yamada2012} and \citet{kulas2012}, respectively. Both studies find that multiple-peaked \lya\ line profiles comprise $\gtrsim$50\% of their respective samples, which should be regarded as lower limits due to the limited $S/N$ and spectral resolution of the \lya\ data. 

Combined with the recent observational evidence that shows that LAEs have smaller \lya\ velocity offsets as compared to LBGs \citep{hashimoto2012}, the spectral substructure that studies such as \citet{yamada2012} and \citet{kulas2012} have uncovered motivates observations of \lya\ at higher spectral resolution. Such observations will probe the ISM and circumgalactic medium through which the \lya\ photons traverse, and provide more stringent constraints on existing \lya\ radiative transfer models. In this paper, we present new higher resolution (120 km s$^{-1}$ FWHM) optical spectra of the \lya\ emission of three galaxies that were initially discovered in the HETDEX Pilot Survey (HPS; \citealp{adams2011}). We discuss these data in $\S$\ref{sec:Data}. In $\S$\ref{sec:Results}, we present basic observational results from our optical spectroscopy and compare the three \lya\ emission line profiles with those of other \lya\ and UV continuum-selected samples. In $\S$\ref{sec:Models}, we present a quantitative comparison of the observed \lya\ line profiles with the predictions of existing \lya\ radiative transfer models with a spherical expanding shell gas geometry. In $\S$\ref{sec:Discussion}, we discuss our findings and finally present our conclusions in $\S$\ref{sec:Conclusions}. 

Throughout this paper, we assume a flat $\Lambda$CDM cosmology ($H_{0} = 70$ km s$^{-1}$ Mpc$^{-1}$, $\Omega_{m} = 0.27$, $\Omega_{\Lambda} = 0.73$; \citealp{komatsu2011}). For the atomic transitions discussed, we use the following vacuum wavelengths in \AA\ quoted from the Atomic Line List v2.04\footnote[7]{Atomic Line List v2.04: \textit{http://www.pa.uky.edu/~peter/atomic/}}: \lya\ (1215.670); \oiii\ $\lambda5007$ (5008.240); \ha\ (6564.610). All magnitudes are reported in the AB system \citep{oke1983}.

\section{OBSERVATIONS AND DATA REDUCTION}\label{sec:Data}

\subsection{Sample Selection - The HETDEX Pilot Survey}\label{subsec:Sample}
The Hobby-Eberly Telescope Dark Energy Experiment (HETDEX; \citealp{hill2008a}) will use 0.8 million LAEs as tracers of the galaxy power spectrum and measure the evolution of dark energy from $1.9 < \z < 3.5$ with high precision. This $\sim$420 deg$^2$ ($\sim$9 Gpc$^3$) blind spectroscopic survey will be conducted with the Hobby-Eberly Telescope and a revolutionary new multiplexed instrument called the Visible Integral-field Replicable Unit Spectrograph (VIRUS; \citealp{hill2012}). As a test-bed for HETDEX, a single prototype VIRUS unit integral field spectrograph (the Mitchell Spectrograph; formerly known as VIRUS-P; \citealp{hill2008b}) has resided at the McDonald Observatory's Harlan J. Smith 2.7 m telescope since 2007 and has conducted a $\sim$100 night, $\sim$169 arcmin$^2$ ($\sim$10$^6$ Mpc$^3$) pilot survey for HETDEX in which 99 star-forming (i.e., non-AGN) LAEs were discovered at redshifts $2<\z<4$. The survey design, catalog, and initial science results can be found in \citet{adams2011} and \cite{blanc2011}.

As described in $\S$\ref{sec:Introduction}, observations of the rest-frame optical nebular emission lines are especially important for constraining models of \lya\ radiative transfer and gas kinematics in star-forming galaxies. Given the aforementioned focus of this paper, we require prior knowledge of the galaxy systemic redshift as measured from the rest-frame optical nebular emission lines. For two HPS LAEs (catalog ID 194 and 256), these data have been published in \citet{finkelstein2011}. Additionally, the only existing optical spectra previous to this work for these galaxies are from the HPS in which the Mitchell Spectrograph observed with 5.5 \AA\ resolution ($\sim$400 km s$^{-1}$ FWHM at 4100 \AA), yielding unresolved \lya\ emission lines \citep{adams2011}. Given the existing data available for these two galaxies, we have chosen to conduct follow-up observations with higher resolution optical spectroscopy. For convenience, some relevant properties of HPS194 and HPS256 are tabulated from the current HPS publications in Table \ref{table:GalData}. The only preselection for the galaxies in this follow-up study is that they have a \lya\ flux $F_{\mathrm{Ly}\alpha} \gtrsim 10^{-16}$ \fluxcgs\ to ease the detection of the rest-frame optical lines (note that 86 of the 99 non-AGN HPS LAEs meet such a selection criteria). Additionally, note that the galaxies included in this study are bright enough in the continuum to be considered LBGs in most surveys (e.g., \citealp{steidel2004}).  

\subsection{Mitchell Spectrograph Optical Spectroscopy}\label{subsec:VPobs}
We obtained follow-up optical spectra of the \lya\ emission of HPS194 and HPS256 using the Mitchell Spectrograph \citep{hill2008b} on the Harlan J. Smith 2.7 m telescope at the McDonald Observatory on the nights of UT 2010 February 13 and 14, respectively. In an attempt to spectrally resolve the \lya\ feature of these galaxies, we replaced the 831 line mm$^{-1}$ grating that was used to carry out the HPS with a 2400 line mm$^{-1}$ grating. This grating yields a spectral resolution of $\Delta\lambda = 1.6$ \AA\ (120 km s$^{-1}$ FWHM at 4100 \AA) over the wavelength range $3690<\lambda$ (\AA) $<4400$ ($R = \lambda/\Delta\lambda \approx 2500$). 

	\begin{figure*}[t]
	\begin{center}
	\begin{tabular}{c}
	\includegraphics[width=17cm]{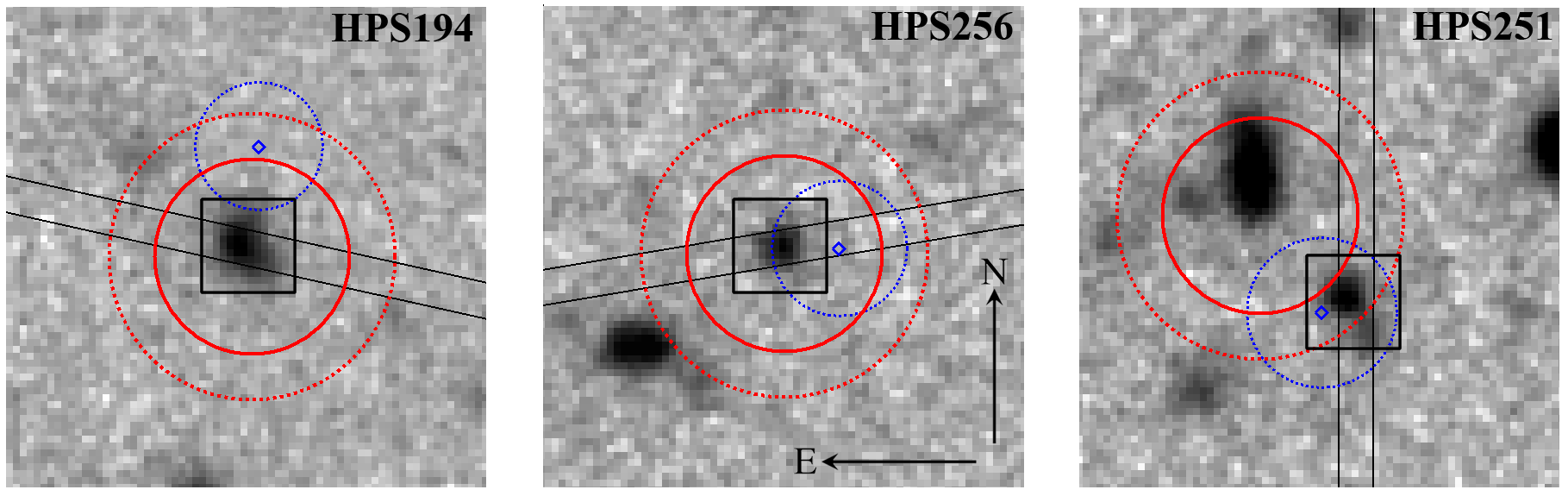}
	\end{tabular}
	\end{center}
	\caption[example] 
	{ \label{fig:PointingThumbs} 
	Subaru $V$-band images ($10.5\arcsec\times10.5\arcsec$ in size) showing the field surrounding each targeted galaxy. The solid red circle indicates the position of the $4.24\arcsec$ diameter Mitchell Spectrograph fiber while the concentric dashed circle indicates the extent of the RMS uncertainty in its position on the sky. The blue diamond and the surrounding dashed blue circle indicate the \lya\ centroid and positional error, respectively, from the HPS \citep{adams2011}. The black lines indicate the width and position angle of the NIRSPEC slit used by \citet{finkelstein2011} for HPS194 and HPS256 and by Song et al. (2013, in preparation) for HPS251. The $2.5\arcsec\times2.5\arcsec$ box is centered on the coordinates of the most likely continuum counterpart to the \lya\ emission \citep{adams2011} and shows the spatial extent of the \textit{HST} images discussed in $\S$\ref{subsec:Imaging}.} 
	\end{figure*} 

The Mitchell Spectrograph is an integral field spectrograph that has high throughput for blue wavelengths ($\sim40$\% at 4100 \AA; \citealp{hill2008b}). The $107\arcsec \times 107\arcsec$ square field of view integral field unit (IFU) contains 246 fibers arranged in a hexagonal close pack pattern with one-third fill factor. Each fiber has a diameter of $4.24\arcsec$ on the sky ($\sim$36 kpc at $z\sim2.4$). For these observations, we position a single fiber on the coordinates of each targeted galaxy that correspond the the most likely optical continuum counterpart to the detected \lya\ emission, as determined by \citet{adams2011}. For the HPS256 pointing, the IFU was positioned such that another nearby HPS LAE (catalog ID 251) fell within the field of view and near the edge of another fiber. Note that HPS194, HPS256, and HPS251 are unresolved within the fiber in the $\lesssim$2$\arcsec$ FWHM seeing during our observations. The remaining fibers in the IFU sample the sky to provide excellent sky subtraction. The astrometry of the individual fiber positions was calibrated to 1.0$\arcsec$ RMS. In Figure \ref{fig:PointingThumbs}, we show Subaru $V$-band images \citep{taniguchi2007} from the Cosmic Evolution Survey (COSMOS; \citealp{scoville2007}; see also $\S$\ref{subsubsec:COSMOS}) in which we indicate the fiber positions relative to the continuum counterparts and the centroid of the \lya\ emission (as determined by \citealp{adams2011}).

The data were taken with $2\times1$ CCD binning in the spectral direction. This results in a dispersion of 0.69 \AA\ per binned pixel (i.e., there are 2.3 binned pixels per resolution element along the spectral direction). The read noise of the CCD is 3.8 $e^{-}$ which allows for sky noise dominated spectra in the 1800 second integration times used throughout the observing run. The CCD gain was set to 1.0 $e^{-}$ ADU$^{-1}$.  For both the HPS194 and the HPS256 pointings, we obtained six hours of total integration in dark and clear conditions. All science exposures were acquired at an airmass of $<1.2$. Two flux standard stars were observed each night to correct for the instrumental response.   

	\begin{figure*}[t]
	\begin{center}
	\begin{tabular}{c}
	\includegraphics[width=17cm]{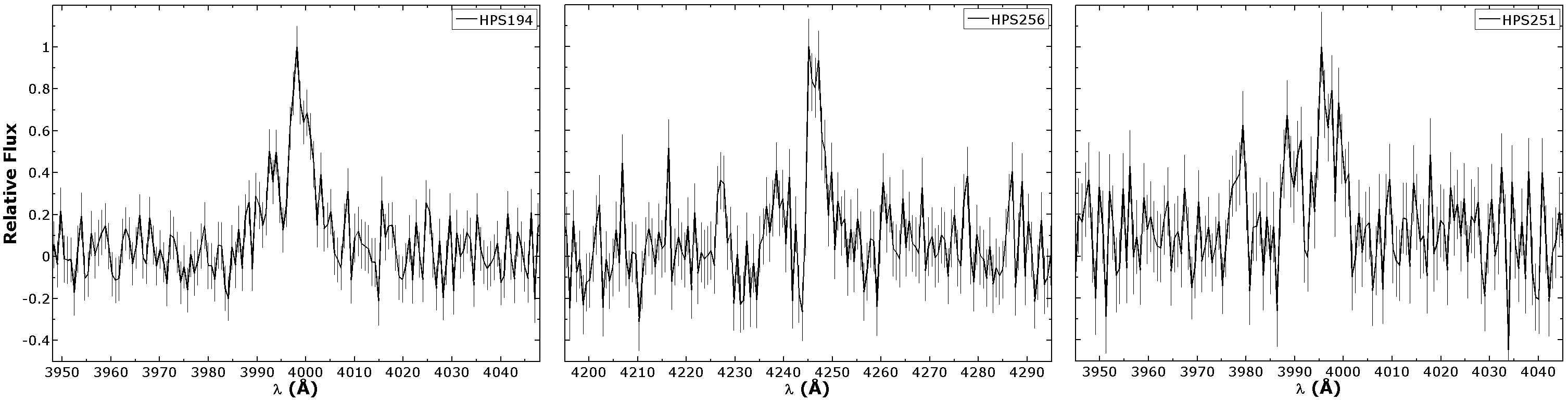}
	\end{tabular}
	\end{center}
	\caption[example] 
	{ \label{fig:CalSpec} 
	Mitchell Spectrograph optical spectra of the targeted galaxies plotted against the observed wavelength at $R\approx2500$. The spectra show a $\pm$50 \AA\ window, centered on the detected \lya\ feature. Each spectrum has been normalized to the peak flux observed in the line. The error bars indicate the $1\sigma$ statistical uncertainties at each wavelength, which are typically $\pm0.11$ for HPS194, $\pm0.14$ for HPS256, and $\pm0.17$ for HPS251 in the relative flux units displayed.}
	\end{figure*} 

The data were reduced with the custom pipeline VACCINE \citep{adams2011}. VACCINE carries out standard reduction procedures for overscan and bias correction. Twilight flats, which were taken both at dawn and dusk each night, are combined and used to locate and trace each of the 246 spectra on the CCD. To properly correct the pixel-to-pixel variations, fiber-to-fiber relative transmission, and the fiber spatial profile, the solar spectrum must be removed from the twilight flats which requires a wavelength solution. Unfortunately, the suite of calibration lamps available did not yield enough comparison lines for determining a reliable wavelength solution for our instrumental setup. Therefore, we extracted a wavelength solution from the twilight flats themselves by convolving a template solar spectrum \citep{kurucz1984} to the instrumental resolution and fitting the result to the combined twilight flats. The fit for each fiber is parameterized as a polynomial of the fourth degree as a function of pixel, where the second order terms are constrained across the fibers to be a smooth function for greater solution stability. An independent wavelength solution results for each fiber on each night. From the RMS in the residuals of the fitted solar features, we expect this wavelength solution to be accurate to $\sim$0.1 \AA\ ($\sim7$ km s$^{-1}$ for \lya\ at $z\sim2.4$). By measuring the centroids of solar absorption features in the dusk and dawn twilight flats separately (which were taken at the high and low temperatures recorded each night, respectively), we find that this wavelength solution systematically drifts with changing thermal conditions by $<0.15$ \AA\ ($<11$ km s$^{-1}$ for \lya\ at $z\sim2.4$). Once the spectra for a given night are combined (see below), this systematic error results in a negligible degradation of the spectral resolution of the final reduced spectrum. After correcting the combined twilight flat for the solar spectrum and normalizing, the science data are extracted and flat fielded. The combined sky spectrum from the empty fibers is fit with a B-spline and subtracted from the fiber containing the target galaxy \citep{dierckx1993,kelson2003}. Finally, cosmic rays are masked. The resulting individual science spectra are collapsed along the fiber spatial direction by a weighted mean ignoring the pixels that were masked for cosmic rays, where the weights are determined from the Poisson and Gaussian uncertainties due to photon counting and read noise, respectively.

Each 1-D science spectrum is further corrected for atmospheric extinction using a model specific to McDonald Observatory as well as for the instrumental response using the coadded flux standard star spectra. The individual corrected 1-D science spectra for each galaxy are then combined by a weighted, $\pm$3$\sigma$-clipped mean, where the weight for each spectrum is determined from differential photometry of stars that were congruently observed with the science exposures by the Mitchell Spectrograph's CCD guider. Due to the stability of the instrument (see the previous paragraph), a single wavelength solution applies for a given fiber throughout the entire night and no resampling of the science spectra are required before combination. Upon examination of the combined spectra for each galaxy (including HPS251), we find that we have made detections of the \lya\ feature in which multiple emission components are visible at $S/N \gtrsim 3$. Using an assumed wavelength-independent index of refraction for air at the altitude of McDonald Observatory ($n_{\mathrm{atm}} = 1.00022$), each combined 1-D spectrum is shifted into vacuum conditions. Finally, for proper comparison with the rest-frame optical nebular emission lines (see $\S$\ref{subsubsec:NIRSPEC}), we further correct the wavelength scale for the relative motion of the Earth with respect to the local standard of rest at the time of observation using the $V_{LSR}$ calculator.\footnote[8]{$V_{LSR}$ Calculator, based on Chapter 6.1 of \citet{meeks1976}: \textit{http://www.astro.virginia.edu/emm8x/utils/vlsr.html}} In Figure \ref{fig:CalSpec}, we show a $\pm$50 \AA\ subsection of the final 1-D spectra, centered on \lya. Since we have successfully detected the \lya\ emission from HPS251 despite it not being a primary target, we include a compilation of its observed properties in Table \ref{table:GalData} along with HPS194 and HPS256. 

\subsection{Ancillary Data}\label{subsec:Ancillary}

\subsubsection{NIRSPEC Rest-Frame Optical Spectroscopy}\label{subsubsec:NIRSPEC}
The interpretation of our \lya\ spectra relies on knowing the galaxies' systemic redshifts $z_{\mathrm{sys}}$. The data required for making this measurement were taken at the 10 m Keck II telescope with the NIRSPEC instrument \citep{mclean1998}, as presented by \citet{finkelstein2011} for HPS194 and HPS256 and Song et al. (2013, in preparation) for HPS251 (also, see \citealp{song2013}). These authors have detected at least one rest-frame optical nebular emission line for each galaxy at $>3\sigma$ significance (\ha\ for all three galaxies in addition to \oiii\ $\lambda$5007 for HPS194 and HPS256) and used the observed wavelengths to calculate $z_{\mathrm{sys}}$. However, \citet{finkelstein2011} did not use the vacuum wavelengths of the \oiii\ $\lambda$5007 and \ha\ transitions when comparing to the corresponding observed wavelengths (which \textit{were} corrected into a vacuum frame and to the local standard of rest). Song et al. (2013, in preparation) have performed a new and improved reduction of the \citet{finkelstein2011} HPS194 and HPS256 data and have recalculated the \oiii\ $\lambda$5007 and \ha-based redshifts using the vacuum wavelengths listed in $\S$\ref{sec:Introduction}. The newly calculated weighted average values of $z_{\mathrm{sys}}$ for these galaxies are listed in Table \ref{table:GalData} along with each measurement's statistical uncertainty. The 1-D \ha\ emission line profile for each galaxy can be seen in lower panels of Figure \ref{fig:VSpec}. 

As described by \citet{finkelstein2011}, small errors in the wavelength solutions of the $H$ and $K$-band NIRSPEC data (i.e., the instrumental setups for detecting \oiii\ $\lambda$5007 and \ha, respectively) result in slightly different values of $z_{\mathrm{sys}}$ for a given galaxy depending on which emission line is being measured. As a result, \citet{finkelstein2011} assigned an additional systematic error term to $z_{\mathrm{sys}}$. After the new reduction by Song et al. (2013, in preparation), the $z_{\mathrm{sys}}$ systematic error for HPS194 and HPS256 has been improved to $\pm0.00013$ and $\pm0.00014$, respectively ($\sim12$ km s$^{-1}$ at $z\sim2.4$). Note that the systematic error cannot be calculated for HPS251 as described by \citet{finkelstein2011} since \ha\ was the only line used to determine its $z_{\mathrm{sys}}$. To be conservative in our analysis, we will assume that the larger systematic error associated with the HPS256 $z_{\mathrm{sys}}$ measurement also applies for HPS251.

\subsubsection{COSMOS \& CANDELS Public Data}\label{subsubsec:COSMOS}
All three galaxies reside within the COSMOS\footnote[9]{COSMOS Archive - Released Datasets:\\ \textit{http://irsa.ipac.caltech.edu/data/COSMOS/datasets.html}} survey \citep{scoville2007} and Cosmic Assembly Near-infrared Deep Extragalactic Legacy Survey (CANDELS; \citealp{grogin2011}) footprints. When necessary in this work, we utilize the wealth of publicly available data from these surveys and note the relevant catalog(s) and citation(s). In particular, we will utilize the CANDELS version 1.0 \textit{Hubble Space Telescope} (\textit{HST}) Advanced Camera for Surveys (ACS) Wide Field Channel (WFC) F606W images and Wide Field Camera 3 (WFC3) / Infrared (IR) channel F160W images to consider the spatial morphologies and distribution of sources around the galaxies \citep{koekemoer2011}. The former probes the three galaxies in the rest-frame UV at $\sim$1800 \AA\ while the latter does so in the rest-frame optical at $\sim$4700 \AA.

\section{OBSERVATIONAL RESULTS}\label{sec:Results}

	\begin{figure*}[t]
	\begin{center}
	\begin{tabular}{c}
	\includegraphics[width=18cm]{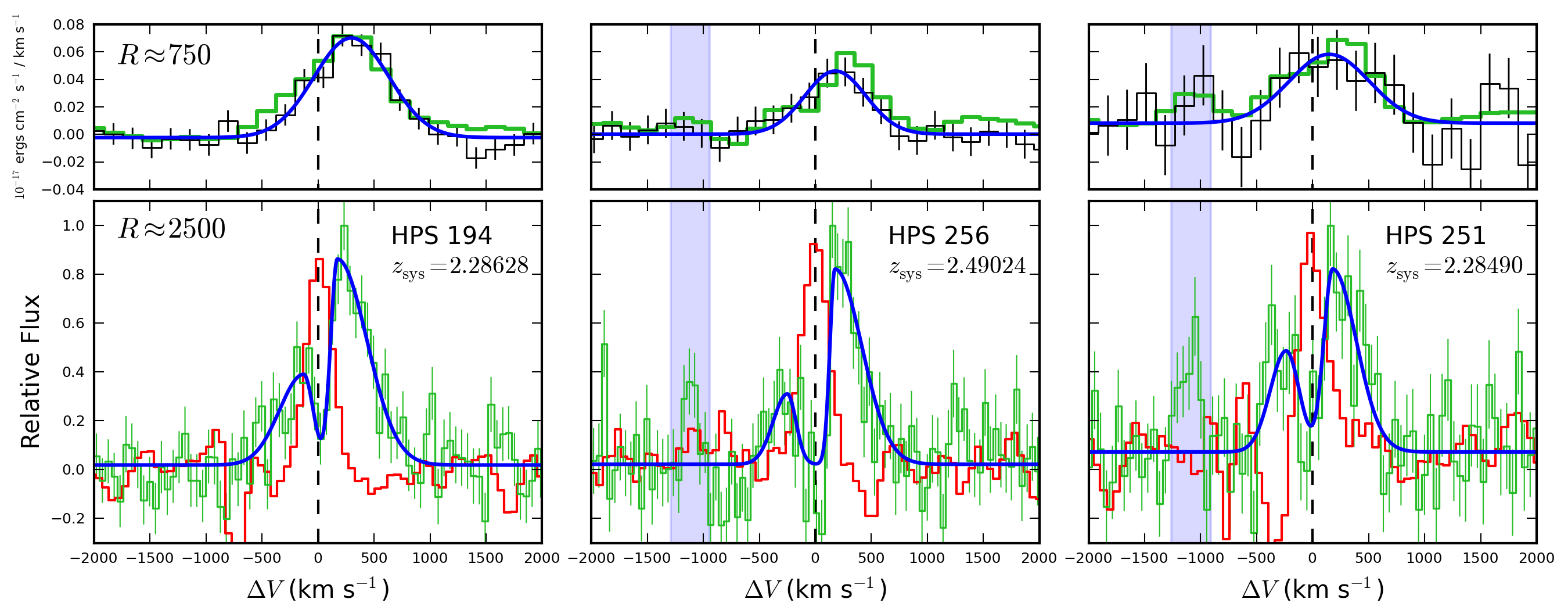}
	\end{tabular}
	\end{center}
	\caption[example] 
	{ \label{fig:VSpec} 
	The \lya\ spectra of the three galaxies shown in velocity space. In the lower panels, we show the $R\approx2500$ \lya\ spectra as the thin green histogram. The blue curves in these panels represent the multi-component asymmetric Gaussian fits, as discussed in $\S$\ref{subsec:LyAFitting}. The red histogram shows the \ha\ spectra for each object from Song et al. (2013, in preparation). The top panels show the coadded $R\approx750$ HPS spectra for each galaxy in black along with Gaussian fits to the emission lines in blue. For comparison, we also show the $R\approx2500$ \lya\ spectra after being degraded to the HPS spectral resolution in green. The light-blue shaded region in the HPS256 and HPS251 panels indicates the location of the second blueshifted peak examined in $\S$\ref{subsubsec:LyAnPeaks}.}
	\end{figure*} 

\subsection{\lya\ Velocity Offsets from the HETDEX Pilot Survey}\label{subsec:HPSoffset}
\citet{finkelstein2011} determined the \lya\ velocity offset from the systemic line center $\Delta v_{\mathrm{Ly}\alpha}$ by comparing the redshifts as determined from the rest-frame optical lines ($z_{\mathrm{sys}}$) and that for \lya\ ($z_{\mathrm{Ly}\alpha}$). They originally found $\Delta v_{\mathrm{Ly}\alpha} = 162 \pm 37 \pm 42$ km s$^{-1}$ and $36 \pm 35 \pm 18$ km s$^{-1}$ for HPS194 and HPS256, respectively. The first set of quoted uncertainties are statistical and display the uncertainty in locating the centroid of a low $S/N$ emission line while the second set are systematic resulting from the differences in the wavelength solutions of the $H$ and $K$-band NIRSPEC data (see \citealp{finkelstein2011} for details). The systematic error resulting from the wavelength calibration of the HPS data is insignificant compared to these error terms \citep{adams2011}. Given the corrected $z_{\mathrm{sys}}$ values discussed in $\S$\ref{subsubsec:NIRSPEC}, these velocity offsets are actually larger than what was originally calculated by \citet{finkelstein2011}. 

During the reduction of our $R\approx2500$ \lya\ spectra, we compared the central \lya\ wavelength to that quoted for each respective object in the HPS catalog \citep{adams2011}. Through this comparison, we discovered that the central wavelengths quoted in Table 3 of that work are actually in error, which was caused by a division of the observed wavelengths by the atmospheric index of refraction $n_{\mathrm{atm}}$ (rather than a multiplication) when correcting to vacuum conditions. To recover the central \lya\ wavelengths with a proper correction into vacuum conditions, one can multiply the values of column 4 in that table by $n_{\mathrm{atm}}$$^{2} = (1.00022)^{2}$. This corresponds to a 0.9 \AA\ (65 km s$^{-1}$) increase in the original quoted central wavelengths for \lya\ at $z \sim 2.4$. With the new corrected values of $z_{\mathrm{sys}}$, applying this additional correction as well as a $V_{LSR}$ correction to the HPS optical data of \citet{adams2011} results in new velocity offsets of $\Delta v_{\mathrm{Ly}\alpha} = 303$ km s$^{-1}$ and $177$ km s$^{-1}$ for HPS194 and HPS256, respectively. In addition, we reevaluate the statistical uncertainty in the velocity centroids by fitting a Gaussian to $10^{3}$ Monte Carlo realizations of the observed HPS spectrum, where the flux in each realization was varied according to the derived $1\sigma$ statistical errors (assuming they are normally distributed). The uncertainties encompass the 68\% confidence interval, and are $\pm28$ km s$^{-1}$ for HPS194 and $^{+52}_{-68}$ km s$^{-1}$ for HPS256. We apply the same corrections and uncertainty estimation methods to the HPS \lya\ data for HPS251; using the NIR results of Song et al. (2013, in preparation), we calculate $\Delta v_{\mathrm{Ly}\alpha} = 146^{+116}_{-156}$ km s$^{-1}$. These results are tabulated in Table \ref{table:GalData}.

In the top panels of Figure \ref{fig:VSpec}, we show the coadded $R\approx750$ HPS spectra (using data from all HPS fibers in which \lya\ was detected). Two of the three galaxies are shown to have statistically significant velocity offsets, which may suggest the presence of large-scale outflows in their ISM. It will be seen when examining the higher resolution $R\approx2500$ spectra that the bulk offset of the unresolved emission lines at low spectral resolution are difficult to interpret due to the complicated \lya\ radiative transfer in these galaxies.

\subsection{Characterization of the $R \approx 2500$ \lya\ Spectra}\label{subsec:LyAFitting}
\subsubsection{Multiple-Peaked \lya\ Emission Line Profiles}\label{subsubsec:LyAnPeaks}
In the lower panels of Figure \ref{fig:VSpec}, we use $z_\mathrm{sys}$ for each galaxy to convert the wavelength scale of the $R\approx2500$ spectra into velocity space, where $\Delta v = 0$ km s$^{-1}$ corresponds to the \ha\ line center. Each galaxy appears to display a complex \lya\ line profile with at least two emission components straddling the velocity zero-point with absorption (i.e., lack of emission) at the \lya\ line center. In all cases, the strongest emission component lies redward of the velocity zero-point and appears quite asymmetric with an extended redward tail. Additionally, each galaxy shows significant emission blueward of the \lya\ line center. HPS194 shows a single, relatively weak blueward emission peak. HPS256 and HPS251 show a similar weak blue peak as HPS194 \textit{in addition to} second blueshifted peaks located at approximately $-1000$ km s$^{-1}$ for each galaxy with $S/N = 2.8$ and $3.0$, respectively.

Recent results have shown that multiple-peaked \lya\ emission line profiles are common among star-forming galaxies (e.g., \citealp{kulas2012,yamada2012}), especially the ``characteristic'' double-peaked profile having a stronger red peak with an accompanying weaker blue emission component straddling the velocity zero-point. The \lya\ line profile of HPS194 easily fits this description. \citet{kulas2012} find that such profiles (``Group 1'' in their nomenclature) consist of 61\% of their sample of 18 $z \sim 2-3$ \lya\ emitting LBGs that were preselected to have multiple \lya\ peaks with measures of $z_\mathrm{sys}$ from observations of \ha\ or \oiii\ in the NIR. Additionally, while \citet{yamada2012} have no measure of $z_{\mathrm{sys}}$ for their large sample of 91 $z\sim3.1$ LAEs in the SSA22-Sb1 field without a multiple-peaked preselection, they do find that $\geq44$\% of their measured \lya\ spectra appear ``Group 1''-like in having two emission components where the redder component is stronger. We present further discussion and more examples of double-peaked \lya\ line profiles from the literature in $\S$\ref{subsec:Comparison}. 

Excluding the highly blueshifted emission components at $-1000$ km s$^{-1}$ in the HPS256 and HPS251 \lya\ spectra, these two galaxies also appear to fit the ``Group 1'' \lya\ line profile morphology. As such, we look further into the nature of the bluest peak for each of these two galaxies. The peaks under scrutiny are marked in Figure \ref{fig:VSpec} by the light-blue shaded regions, which have been centered on the central velocity of the peak as determined through a Gaussian fit to the emission component in the $R\approx2500$ data. The width of the shaded region corresponds to one spectral resolution element of the $R\approx750$ data. Since the velocity offset of this emission peak is so large, it should be cleanly resolved in the HPS $R\approx750$ spectra. However, we see no evidence of emission beyond the 1$\sigma$ statistical uncertainties at the corresponding velocity in the $R\approx750$ data for either HPS256 or HPS251. Note that this does not rule out the reality of these features in the $R\approx2500$ spectra. In the top panels of Figure \ref{fig:VSpec}, we also show the $R\approx2500$ data convolved to the spectral resolution and interpolated to the dispersion of the HPS. As can be seen, the lower spectral resolution smears out the weak emission peak to within the 1$\sigma$ statistical uncertainties for both galaxies, so we do not expect to clearly detect them in the HPS spectra.

We have examined the possibility that the low $S/N$ second blueward peaks are the result of systematic or random effects in the $R\approx2500$ data. Despite that HPS256 and HPS251 were observed in the same field, the two galaxies have different $z_{\mathrm{sys}}$. Thus, the second blueshifted peaks do not correspond to the same observed wavelength even though they have very similar velocity offsets relative to the systemic redshift. We have examined the extracted sky spectra from the fibers in the IFU that are adjacent to those containing the LAEs to look for sky absorption or emission features in the vicinity of the second blueward peaks. We find none for HPS251. For HPS256, the second blueward emission peak lies within one spectral resolution element of a sky absorption feature. However, we find no evidence of a systematic sky subtraction error since we do not see peaks with similar $S/N$ in the reduced spectra of adjacent fibers at the wavelength corresponding to that of the second blueward emission peak. Additionally, we have visually examined the 2-D spectra of the fibers containing the LAEs in each individual CCD exposure after the cosmic ray masking procedure to verify that no cosmic rays or other artifacts were left unmasked in the vicinity of the \lya\ spectra. Finally, we experimented with our data combination rejection procedure by recomputing the $\pm$3$\sigma$-clipped weighted mean on the 2-D data before collapsing to the final 1-D spectra. Performing the combination with the increased pixels space should yield a more robust rejection of outlying pixel values. After collapsing this combined 2-D spectra along the fiber spatial direction, we find that the resulting reduced \lya\ line profiles are statistically identical to those in the original data reduction described in $\S$\ref{subsec:VPobs} and that the second blueward peaks are insensitive to the specifics of the data combination procedure. Thus, we do not believe that the peaks at $-1000$ km s$^{-1}$ for HPS256 and HPS251 are the result of systematic or random effects in the $R\approx2500$ data. 

We look to the distribution of the sources on the sky within the extent of the fiber by examining the COSMOS $V$-band images shown in Figure \ref{fig:PointingThumbs}. Both of our primary targeted galaxies (i.e., HPS194 and HPS256) are centered within the fiber with no additional continuum sources located within the RMS positional error. The case for HPS251 is more complicated. Since it was not originally a primary target of our observations, the location of the LAE continuum counterpart is not positioned optimally with respect to the fiber center. As a result, the area of sky covered by the fiber contains several additional continuum sources. These include a brighter and extended source with a photometric redshift\footnote[10]{COSMOS Photometric Redshift Catalog, November 2008 Release. The quoted uncertainties correspond to the 99\% confidence interval.} of $0.97_{-0.05}^{+0.04}$ \citep{ilbert2009}. If the second HPS251 blue peak was from this lower-redshift galaxy, its rest-frame wavelength would be $2019_{-40}^{+53}$ \AA. Since this range does not correspond to any prominant near-UV transition, the low-redshift galaxy in the fiber is most likely not the source of the second blue peak\footnote[11]{The photometric redshift of this brighter and extended source is confirmed with a detection of \ha\ at 12989 \AA\ ($z=0.98\pm0.01$) in the 3D-HST survey \citep{brammer2012}.}. The remaining source other than the LAE itself within the fiber (taking into account the RMS uncertainty in its position) is a faint source to the east of the low-redshift galaxy. This source has no COSMOS photometric redshift. 

Since we are unable to definitively determine another source from which they originate, we cannot rule out that the second blueward peaks are indeed \lya\ emission from the targeted LAEs. A triple-peaked \lya\ line displaying two weak blue peaks has not been previously observed in studies that looked for multiple-peaked \lya\ emission (e.g., \citealp{kulas2012,yamada2012}). Additionally, there are no clear examples of such highly blueshifted \lya\ emission, even in the most extreme starburst LAEs in the nearby universe when observed with high resolution and sensitivity (see \citealp{heckman2011}). With the inconclusive nature of the second blue peaks and given the prevalence of ``Group 1''-type double-peaked profiles in the literature, we continue forward treating HPS256 and HPS251 as double-peaked objects (i.e., we ignore the second blueward peaks). However, we will return to the possibility of these galaxies being more complex, triple-peaked \lya\ systems in $\S$\ref{sec:Discussion}.

\subsubsection{Quantitative Description of the \lya\ Emission}\label{subsubsec:LyAQuantitative}
To quantitatively characterize the \lya\ line profiles and extract observable quantities from the double-peaked profiles, we utilize the MPFIT IDL package \citep{markwardt2008} to fit a function $f_{\mathrm{tot}}(\Delta v)$ to each galaxy's \lya\ spectrum in velocity space with the following form:
\begin{equation}\label{eq:flux}
f_{\mathrm{tot}}(\Delta v) = f_{blue}(\Delta v)\; +\; f_{red}(\Delta v)\; +\; C \; ,
\end{equation}
where $C$ is the continuum level and $f_{x}(\Delta v)$ is a function describing each component of emission (e.g., $f_{x}$ denotes the red emission component if $x = red$ and the blue component if $x = blue$). At the spectral resolution of our data, the emission components are asymmetric. For $f_{x}(\Delta v)$, we thus adopt an ``asymmetric Gaussian\footnote[12]{The ``asymmetric Gaussian'' function is implemented in our IDL code through the \texttt{ARM\_ASYMGAUSS} routine developed by Andrew Marble: \textit{http://hubble.as.arizona.edu/idl/arm/}}'' functional form similar to that used by \citet{mclinden2011} to fit the \lya\ line profiles of $z = 3.1$ LAEs. This is given by:
\begin{equation}\label{eq:componentflux}
f_{x}(\Delta v) = A_{x}\; \exp\left[-\frac{(\Delta v - \Delta v_{0,x})^2}{2 \sigma_{x}^2}\right] \; ,
\end{equation}
where, $A_{x}$ is the amplitude of the respective emission component above the continuum level, $\Delta v_{0,x}$ is the velocity of the peak emission of the component, and $\sigma_{x}$ is a width parameter. Skew is introduced to the Gaussian described above by defining $\sigma_{x}$ as:
\[\sigma_{x} \equiv \left\{ 
\begin{array}{l l}
  \sigma_{x,b} & \quad \mathrm{if}\; \Delta v < \Delta v_{0,x}\\
  \sigma_{x,r} & \quad \mathrm{if}\; \Delta v > \Delta v_{0,x}\\ \end{array} \right.\; . \]
The FWHM of component $x$ is then given by:
\begin{equation}\label{eq:FWHM}
\mathrm{FWHM}_{x} = \sqrt{2 \ln(2)}\; (\sigma_{x,b} + \sigma_{x,r})\; ,
\end{equation}
and we define a parameter describing the skew as:
\begin{equation}\label{eq:skew}
\alpha_{x} \equiv \sigma_{x,r}\: /\: \sigma_{x,b}\; .
\end{equation}
This parameter is defined such that $\alpha_{x} = 1$ describes a symmetric emission component, while $\alpha_{x} > 1$ ($< 1$) describes an asymmetric emission component with an extended red (blue) tail. We extract a total of seven observables describing the \lya\ line profile morphology: the red-to-blue component flux ratio $F_{red} / F_{blue}$ (where $F_{x} = \int f_{x}(\Delta v)\; d(\Delta v)\:)$, the red component velocity offset from systemic ($\Delta v_{0,red}$), the total velocity separation of the red and blue components ($\Delta v_{\mathrm{tot}} = \Delta v_{0,red} - \Delta v_{0,blue}$), FWHM$_{blue}$, FWHM$_{red}$, $\alpha_{blue}$, and $\alpha_{red}$. We do not include $C$ since the continuum flux for all three galaxies is well below the detection limit of our spectra. Note that the functional form of $f_{\mathrm{tot}}(\Delta v)$ is not motivated by the physics of the \lya\ radiative transfer, but is rather for the purpose of parameterizing the emission line morphology for comparison with previously published results. This fitting procedure will be especially useful in the future for characterizing and finding trends in large samples of observed \lya\ emission line profiles.

In the bottom panels of Figure \ref{fig:VSpec}, we show the best-fit $f_{\mathrm{tot}}(\Delta v)$ to each \lya\ emission line, the parameters of which are listed in Table \ref{table:FitParam}. The quoted statistical uncertainties encompass the 68\% confidence interval and were determined by fitting Equation \ref{eq:flux} to $10^{3}$ Monte Carlo realizations of the observed spectrum, similar to the method described in $\S$\ref{subsec:HPSoffset} for the low resolution HPS \lya\ spectra. In addition to these statistical uncertainties, recall that there also exists a systematic uncertainty in the velocity zero-point resulting from the NIR data (see $\S$\ref{subsubsec:NIRSPEC}; this only affects the $\Delta v_{0,red}$ measurement). All applicable values reported in Table \ref{table:FitParam} (i.e., FWHM$_{x}$ and $\alpha_{x}$) have not been corrected for the instrumental resolution. Such a correction would decrease the FWHM$_{x}$ measurements (by no more than 30 km s$^{-1}$ for the narrowest component) and increase (decrease) the $\alpha_{red}$ ($\alpha_{blue}$) measurements, thus yielding the asymmetry parameterizations as lower (upper) limits. As a result, these limits on $\alpha_{x}$ clearly indicate that the red \lya\ peak for each LAE is quite asymmetric with a red tail, while the blue \lya\ peak tends to be asymmetric with a blue tail.

One may notice that the measured $\Delta v_{0,red}$ given in Table \ref{table:FitParam} may not necessarily agree with the $\Delta v_{\mathrm{ly}\alpha}$ measurements from the lower resolution HPS observations in Table \ref{table:GalData}. This is due to the low $S/N$ of both sets of spectra and the intrinsic asymmetry in the dominant redward \lya\ emission peak (in particular, the extended red tail and truncated blue edge of that component). When observing such an asymmetric emission line at the lower spectral resolution of the HPS, the excess emission in the extended red tail pulls the convolved peak of the unresolved line further redward. This can be seen in the top panel of Figure \ref{fig:VSpec} when comparing the actual HPS $R\approx750$ data with the $R\approx2500$ data that has been degraded to the HPS spectral resolution and dispersion. We discuss this effect further in $\S$\ref{subsec:Correlation?}.

\subsection{Double-Peaked \lya\ Emission Across Galaxy Samples}\label{subsec:Comparison}
The presented sample of three LAEs were \textit{not} preselected to have multiple \lya\ emission components. However, it should not be surprising that each LAE shows a multiple-peaked morphology at high spectral resolution since multiple peaks are a natural outcome of \lya\ resonant line transfer. This is due to photons diffusing through the scattering medium in real and frequency space until reaching the line wings on either side of the core where the optical depth is low enough for escape. For the simplest case of resonant scattering through static gas, the result is a double-peaked emergent spectrum that is symmetric about the velocity zero-point \citep{neufeld1990}. The observed frequency of multiple peaked \lya\ emission appears to be significant for LAE samples ($\sim50$\% in the overdense SSA22-Sb1 field; \citealp{yamada2012}) as well as for LBG samples ($\sim30$\%; \citealp{kulas2012}). These frequencies should be regarded as lower limts due to the limited spectral resolution and $S/N$ of the data. Disregarding possible trends with environment (i.e., overdensity vs. field), these results suggest that drawing three LAEs with multiple peaks from the HPS sample is not an unlikely scenario. 

	\begin{deluxetable}{ccccc}
	\tabletypesize{\scriptsize}
	\tablecaption{Best-Fit $R\approx2500$ \lya\ Observables\label{table:FitParam}}
	\tablehead{ & \colhead{(Units)} & \colhead{HPS194} & \colhead{HPS256} & \colhead{HPS251} \\ }
	\startdata
	$F_{red} / F_{blue}$ & ( - ) & $2.4\pm0.4$ & $3.9^{+0.9}_{-1.2}$ & $2.0^{+0.7}_{-0.4}$\\[0.8ex]
	$\Delta v_{0,red}$ & (km s$^{-1}$) & $173^{+17}_{-20}$ & $175^{+9}_{-19}$ & $179^{+8}_{-28}$\\[0.8ex]
	$\Delta v_{\mathrm{tot}}$ & (km s$^{-1}$) & $300^{+19}_{-71}$ & $425^{+33}_{-38}$ & $415^{+29}_{-44}$\\[0.8ex]
	FWHM$_{blue}$ & (km s$^{-1}$) & $353^{+48}_{-62}$ & $240^{+75}_{-23}$ & $306^{+81}_{-37}$\\[0.8ex]
	FWHM$_{red}$ & (km s$^{-1}$) & $380^{+44}_{-23}$ & $335^{+11}_{-80}$ & $343^{+45}_{-18}$\\[0.8ex]
	$\alpha_{blue}$ & ( - ) & $0.4^{+0.1}_{-0.2}$ & $0.6^{+0.4}_{-0.1}$ & $0.8\pm0.2$\\[0.8ex]
	$\alpha_{red}$ & ( - ) & $4.4^{+1.2}_{-0.9}$ & $5.3^{+1.5}_{-0.5}$ & $2.7^{+0.8}_{-0.1}$\\
	\enddata
	\tablecomments{See $\S$\ref{subsec:LyAFitting} for the definitions of these parameters. The quoted uncertainties are statistical. Systematic errors associated with the optical and NIR wavelength calibrations are discussed in $\S$\ref{subsec:VPobs} and \ref{subsubsec:NIRSPEC}, respectively (these only affect the $\Delta v_{0,red}$ measurement).}
	\end{deluxetable}

At low spectral resolution, the measured velocity offsets of all three LAEs in our sample appear to be consistent with the mean \lya\ velocity offset as measured from 8 other \lya\ selected galaxies in the literature that were observed at similar resolution ($\langle\Delta v_{\mathrm{Ly}\alpha}\rangle = 164 \pm 97$ km s$^{-1}$, where the uncertainty represents the standard deviation of the sample; \citealp{mclinden2011,hashimoto2012,guaita2013}). This value, as noted by \citet{finkelstein2011} and the aforementioned authors, is much smaller than the $\sim450$ km s$^{-1}$ offset that is typically measured for LBGs (e.g., \citealp{steidel2010}).  

From the results of \citet{yamada2012} and \citet{kulas2012}, \lya\ spectra having a strong red component and a weaker blue component are the most common multiple-peaked line profile morphology. Besides being observed in both \lya-selected and UV continuum-selected samples of galaxies, this line profile can be observed for galaxies spanning a variety of redshifts from local LBG analogues \citep{heckman2011} to galaxies beyond $z\sim3$ \citep{tapken2007}. In addition, it is not only observed emerging from individual galaxies (e.g., this work, \citealp{mclinden2011,kulas2012,christensen2012,yamada2012}), but also from the large \lya\ nebulae (i.e., \lya\ blobs; LABs) that are often associated with protoclusters \citep{matsuda2006,yang2011} and radio galaxies \citep{adams2009}. 

\begin{figure}[t]
	\begin{center}
	\begin{tabular}{c}
	\includegraphics[width=8.25cm]{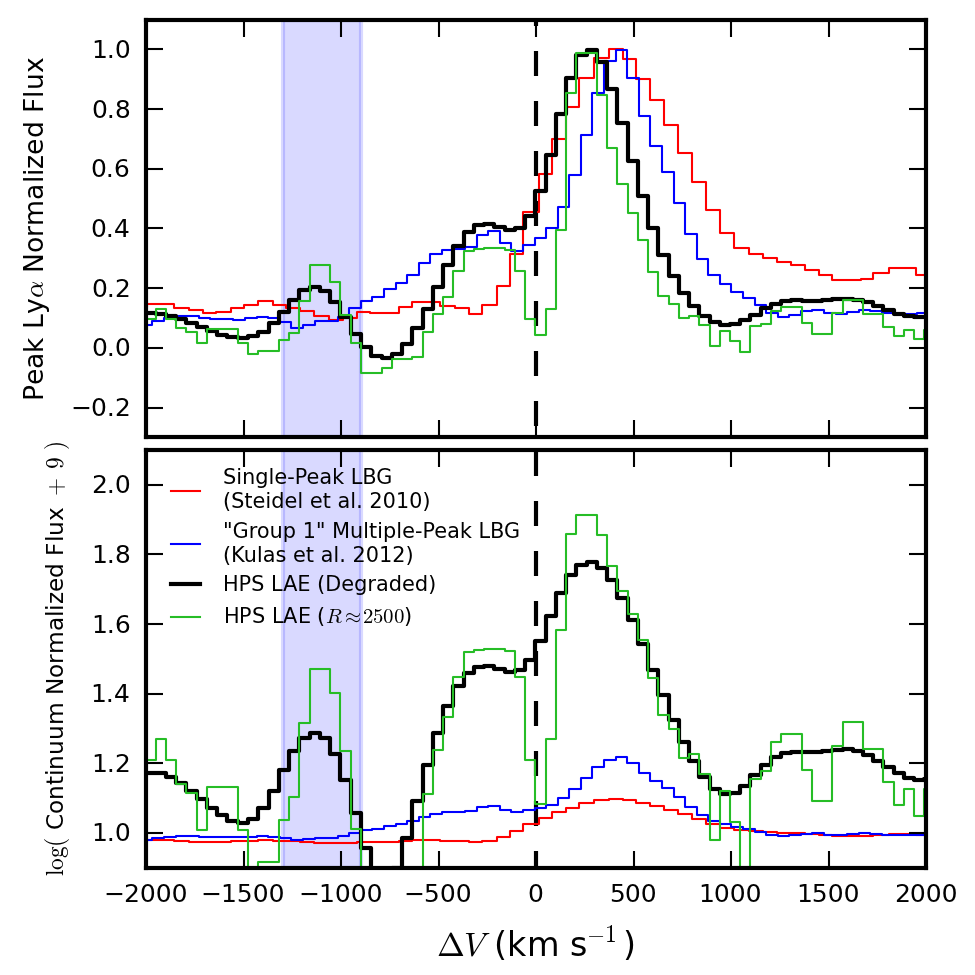}
	\end{tabular}
	\end{center}
	\caption[example] 
	{ \label{fig:K12G1Comparison} 
	Comparison of the \lya\ line profiles for different galaxy samples. In both panels, the blue histogram is the composite \lya\ spectrum for LBGs from \citet{kulas2012} that encompass the ``Group 1'' multiple-peaked profiles and qualitatively match the general \lya\ morphology of the LAEs presented here. The red histogram is the composite spectrum of single-peaked \lya\ profiles of LBGs from \citet{steidel2010}. The green histogram is the average $R\approx2500$ \lya\ line profile for the three HPS LAEs observed in this work, while the black histogram is the same data but degraded to the same spectral resolution as the LBG \lya\ spectra. The top panel is normalized to the peak \lya\ flux to accentuate differences in the velocity axis while the lower panel is normalized to the continuum flux redward of \lya, showing the relative differences in \ewlya. The light-blue shaded region has the same meaning as in Figure \ref{fig:VSpec}.}
	\end{figure}

As with the unresolved \lya\ velocity offsets $\Delta v_{\mathrm{Ly}\alpha}$ discussed above, the velocity offsets of the multiple peaks relative to the systemic velocity might be expected to be different between LAE and LBG samples. From \citet{kulas2012}, the average velocity offset for the red emission component of their ``Group 1'' line profiles is $\langle\Delta v_{0,red}\rangle = 417\pm101$ km s$^{-1}$ while the average total separation between the red and blue components is $\langle\Delta v_{\mathrm{tot}}\rangle = 801\pm136$ km s$^{-1}$. On average for the LAEs presented here, these same measurements are smaller by a factor of $\sim2$ (cf., Table \ref{table:FitParam}). This is visualized in Figure \ref{fig:K12G1Comparison}, which shows the average $R\approx2500$ spectrum of our three LAEs plotted on the same velocity scale relative to $\z_{\mathrm{sys}}$ as the composite spectrum of the 11 ``Group 1'' double-peaked LBGs from \citet{kulas2012}. Additionally, we show a composite spectrum of 29 single peaked LBGs from \citet{steidel2010}. The top panel shows the \lya\ spectra normalized by the peak flux to accentuate differences in the velocity offsets, while the lower panel shows the same spectra normalized to the continuum\footnote[13]{The relative flux of our $R\approx2500$ \lya\ spectra has been scaled to match the total \lya\ line flux measured from the HPS \citep{adams2011}. Since no continuum is detected spectroscopically for our LAEs, we use the continuum flux measured by COSMOS $V$-band photometry for normalization (see Table \ref{table:GalData}).} redward of the \lya\ line to accentuate differences in the relative \ewlya\ for each sample. In both panels, we show a degraded version of our LAE composite \lya\ line profile that has the same average spectral resolution of \citet{steidel2010} and \citet{kulas2012} to show that the differences in the velocity structure between the samples are largely invariant with spectral resolution. This figure embodies a trend that higher \ewlya\ objects tend to have smaller \lya\ velocity offsets (e.g., \citealp{hashimoto2012}). This trend is especially interesting since there is naturally some overlap between LBG and LAE samples; some LAEs, such as those presented in this work, have high \ewlya\ and small \lya\ velocity offsets, but are bright enough in the continuum to be considered LBGs in most surveys and have some physical properties that are similar to the typical LBG, such as stellar mass.

In addition to the three LAEs presented here, the number of \lya\ line profiles displaying the strong red and weaker blue double-peaked morphology with a measure of $z_{\mathrm{sys}}$ has grown significantly (e.g., \citealp{fynbo2010,steidel2010,mclinden2011,yang2011,heckman2011,kulas2012,christensen2012,noterdaeme2012}). As with our three LAEs, the systemic \lya\ line center usually lies in between the two peaks. While the large sample of such \lya\ line profiles from \citet{yamada2012} for LAEs do not have a measure of $z_{\mathrm{sys}}$, these previous results suggest that we can be reasonably safe in at least comparing $\Delta v_{\mathrm{tot}}$ for our galaxies with that from the \citet{yamada2012} sample. Since the \citet{yamada2012} sample has $\langle\Delta v_{\mathrm{tot}}\rangle = 608\pm170$ km s$^{-1}$, our LAEs appear to be on the lower end of the distribution. However, despite the asymmetry that they measure in their \lya\ spectra, \citet{yamada2012} determine the peak-to-peak separation by fitting symmetric Gaussians to each emission component. If we were to fit our \lya\ spectra with symmetric Gaussians consistent with their method, $\Delta v_{\mathrm{tot}}$ would be 422, 561, and 559 km s$^{-1}$ for HPS194, HPS256, and HPS251, respectively. In this case, each of our LAE would be close to within $1\sigma$ of the mean of the \citet{yamada2012} $\Delta v_{\mathrm{tot}}$ distribution. Thus, as also seen at lower resolution, our double-peaked \lya\ spectra appear to be consistent with those already measured for other LAE samples.

Depending on the gas geometry and kinematics, the asymmetric shape of the emission line components can provide clues about the \lya\ radiative transfer and the properties of the scattering gas (e.g., \citealp{verhamme2006,zheng2010,schaerer2011,christensen2012,noterdaeme2012}). As seen from our measurements, both emission peaks in the three LAEs' spectra are typically asymmetric, with a sharper fall-off towards the \lya\ line center. While asymmetry measurements are highly dependent on the instrumental resolution, we can at least compare our \lya\ spectra qualitatively to the asymmetry observed for galaxies showing \lya\ with a similar characteristic double-peaked morphology. At slightly lower spectral resolution than our data (180 km s$^{-1}$ FWHM as compared to our data at 120 km s$^{-1}$), \citet{yamada2012} find that the stronger red peak in objects showing the characteristic double-peaked \lya\ profile is typically asymmetric with an extended red wing on the red emission component. \citet{mclinden2011} also find this for the one object in their sample displaying the double-peaked morphology. At similar or higher spectral resolution to ours, studies such as \citet{fynbo2010}, \citet{heckman2011}, \citet{yang2011}, \citet{noterdaeme2012}, and \citet{christensen2012} have observed objects whose double-peaked \lya\ spectrum is asymmetric in the same manner as our spectra (i.e., each emission component showing asymmetry with the sharpest fall-off towards line center). The asymmetry in our data appear to be consistent with that of the highest resolution \lya\ spectra in the examples given ($\lesssim75$ km s$^{-1}$ FWHM; \citealp{christensen2012}), which shows that the flux fall-off towards line center is extremely sharp. Such asymmetry is thought to be the signature of outflowing gas, where the extended red wing of the red emission component consists of photons that have been ``backscattered'' several times off of the far-side inner surface of an expanding shell of neutral hydrogen gas (e.g., \citealp{verhamme2006}; also, see $\S$\ref{subsec:GridTrends}). In the following section, we examine the ability of such a model to reproduce the shape of the observed \lya\ profiles for our LAEs.

\section{COMPARISON WITH \lya\ RADIATIVE TRANSFER MODELS}\label{sec:Models}
In $\S$\ref{sec:Introduction} and the references therein, the complex processes that govern the escape of \lya\ from a galaxy and their effect on the emergent \lya\ emission line profile were discussed, along with mention of several studies which attempt to model them. To date, the model that observed \lya\ line profiles have been most widely compared with is that of the propagation of \lya\ photons through a simple expanding, spherical shell of neutral hydrogen gas (e.g., \citealp{verhamme2006,barnes2010,schaerer2011}). This is largely due to its success in reproducing many observed properties of $z\sim3$ LBGs, including the shape of the \lya\ line in emission and absorption for various line profile morphologies (e.g., \citealp{schaerer2008,verhamme2008}). While highly idealized, this model is physically motivated by the basic picture of a star-forming galaxy in which the energy from a centrally located starburst (which is the location of the initial \lya\ emission) pushes out the ISM through the combined effect of stellar winds and supernovae explosions. The result is a ``superbubble'' formed within a geometrically thin shell of expanding neutral gas.

\subsection{The Expanding Shell Model Grid}\label{subsec:ModelGrid}
To compare with the $R\approx2500$ data, we have used the \lya\ radiative transfer code developed by \citet{zheng2002} and \citet{kollmeier2010} to produce a grid of model \lya\ spectra resulting from the transfer of \lya\ photons through expanding, isothermal, and homogeneous shells of neutral hydrogen gas. This code was also used by \citet{kulas2012} in their qualitative comparison of model \lya\ line profiles with the multiple-peaked \lya\ spectra of LBGs. The expanding shell radiative transfer models that we run are identical to previous work by \citet{verhamme2006} and \citet{schaerer2011}, with the exception of a couple notable simplifications:

\textit{1) Intrinsic \lya\ Spectrum -} The intrinsic spectrum emitted from the central point source in these two studies is a Gaussian \lya\ line plus a continuum, whereas our models currently only use an idealized monochromatic \lya\ line. The lack of continuum emission in our models results in only negligible effects on the output model \lya\ spectra since we are comparing them with the \lya\ spectra of galaxies that have observed \ewlya\ $>100$ \AA\ in the rest-frame. Using \ha, Song et al. (2013, in preparation) have measured the 1-D integrated line-of-sight nebular velocity dispersion $\sigma_{\mathrm{H}\alpha}$ for each galaxy. This traces the velocity dispersion of the galaxy's \hii\ regions and thus the width of the intrinsic \lya\ emission. For the three LAEs presented here, $\sigma_{\mathrm{H}\alpha}$ is at most 72 km s$^{-1}$ (which is systematically smaller by $\sim40$ km s$^{-1}$ than $\sigma_{\mathrm{H}\alpha}$ for the average $z\sim2$ UV-selected galaxy; \citealp{erb2006}). We discuss the effect of using a dispersion of \lya\ injection frequencies (rather than a monochromatic line) in $\S$\ref{subsec:Limitations}.

	\begin{figure*}[t]
	\begin{center}
	\begin{tabular}{c}
	\includegraphics[width=17cm]{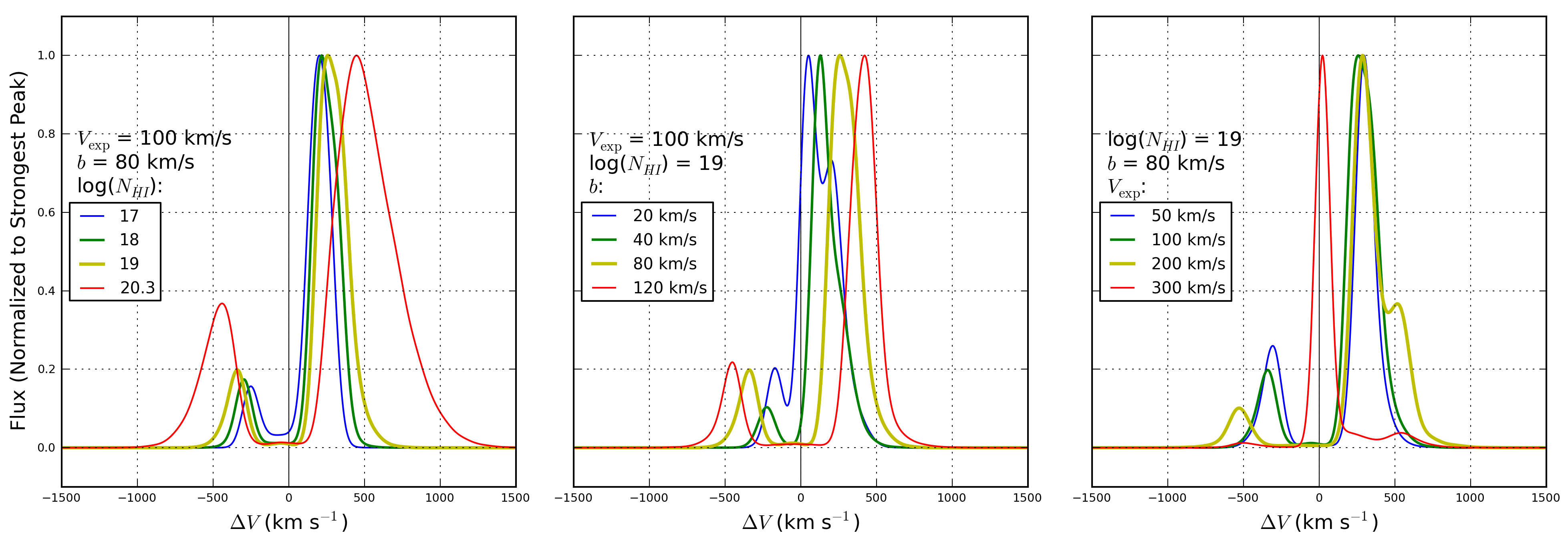}
	\end{tabular}
	\end{center}
	\caption[example] 
	{ \label{fig:ExModelSpec} 
	Representative \lya\ emission line profiles from the grid of \lya\ radiative transfer models in the homogeneous expanding shell geometry. From left to right, each panel shows the variation of the \lya\ emission line profile as functions of \nh, $b$, and \vexp, respectively, while the other two respective parameters are held constant about a fiducial model having \nh\ = $10^{19}$ cm$^{-2}$, $b = 80$ km s$^{-1}$, and \vexp\ = 100 km s$^{-1}$. Each model spectrum has been convolved with a 120 km s$^{-1}$ FWHM Gaussian kernel for comparison with our $R\approx2500$ Mitchell Spectrograph observations of \lya\ and have been scaled to the peak flux.}
	\end{figure*} 

\textit{2) Dusty Shells -} The neutral hydrogen shells in these two studies can be uniformly mixed with dust (parameterized by a dust absorption optical depth $\tau_{a}$, which essentially characterizes the dust-to-gas ratio), whereas we model only dust-free shells. Including dust in the shell has the effect of ``sharpening'' the \lya\ line profile. Basically, this is because photons that emerge farther from the \lya\ line center have encountered more scattering events, which gives them a higher probability of being absorbed by a dust grain. For $\tau_{a} \lesssim 1$, the overall shape of the emergent \lya\ line profile is well preserved when compared to the dust-free line profile emerging from an otherwise identical shell, other than the sharpening effect \citep{verhamme2006,schaerer2011}. The dust absorption optical depth is related to the measured extinction as $E(B-V) \approx 0.1 \tau_{a}$ \citep{verhamme2006}. Using $E(B-V)$ for our three LAEs (cf., Table \ref{table:GalData}; \citealp{blanc2011}), $\tau_{a} \approx 0.9\pm0.6$, $1.0\pm0.9$, and $0.7\pm0.8$ for HPS194, HPS256, and HPS251, respectively. The potential impact of dust on our comparison with dust-free shell models is further discussed in $\S$\ref{subsec:Limitations}.

An individual model on our grid is thus characterized by three parameters that describe the shell: the (uniform) expansion velocity \vexp, the Doppler parameter $b$ (which in the absence of turbulence is equivalent to the gas thermal velocity), and the neutral hydrogen column density \nh. Our grid has been computed for the following values: \vexp\ = 50, 100, 200, and 300 km s$^{-1}$; $b$ = 20, 40, 80, and 120 km s$^{-1}$; \nh\ = $10^{17}$, $10^{18}$, $10^{19}$, and $2\times10^{20}$ cm$^{-2}$. For comparison with the $R\approx2500$ Mitchell Spectrograph \lya\ data, each model \lya\ spectrum has been convolved with a 120 km s$^{-1}$ FWHM Gaussian kernel. 

\subsection{Model Trends}\label{subsec:GridTrends}
\citet{verhamme2006} decompose the \lya\ line profile to determine how various features in multiple-peaked \lya\ spectra arise (see their Figure 12). In summary, there are three basic modes of escape of \lya\ photons from an expanding shell, any number of which could be occurring simultaneously in the same model depending on the shell parameters: \textit{1}) a single series of scatterings before escape; \textit{2}) one or more series of ``backscatterings'' before escape (where a backscattering event is defined as a series of scatterings after which the photon traverses the shell's cavity and reenters the shell in a different location); and \textit{3}) direct escape (i.e., no interaction with the shell). Mode 1, which is a similar process to resonant scattering through a slab of neutral gas, results in a peak blueward of the \lya\ line center \textit{and} a redward peak. The relative strength of these two peaks is determined by \vexp\ (i.e., the two peaks are more equal in strength for small \vexp\ as one approaches the static case). Mode 2 results in \textit{another} peak redward of the \lya\ line center, which is composed of photons that have undergone one backscattering event. Due to radiative transfer effects (i.e., these photons are ``reflected'' off the receding inner surface of the shell back towards the observer), this peak traces $\sim2V_{\mathrm{exp}}$ in terms of its offset from the line center. Photons that undergo more than one backscattering event emerge progressively redward of this peak with less probability, which results in an extended redward tail. Finally, mode 3 occurs only when \vexp\ is large enough for the \lya\ photons emitted at the shell's center to already be redshifted out of resonance in the frame of the expanding gas when they encounter the shell. Therefore, these photons emerge as a peak at the \lya\ line center.

In Fig. \ref{fig:ExModelSpec}, we show several representative model \lya\ spectra from the grid, where each panel shows the variation of one of the three parameters for fixed values of the other two about a fiducial model having \nh\ = $10^{19}$ cm$^{-2}$, $b = 80$ km s$^{-1}$, and \vexp\ = 100 km s$^{-1}$. All of the escape modes discussed above are seen in at least one of the models shown in this figure.  

To understand how the \lya\ line profile that is emergent from the expanding shell depends on the three parameters, we begin by considering the simplest case of resonant scattering through a static slab of neutral hydrogen gas with Doppler parameter $b$ and neutral hydrogen column density \nh. This configuration results in a symmetric double-peaked spectrum about the \lya\ line center (e.g., \citealp{neufeld1990}). Based on a random walk in both frequency space and real space, one can show that the typical velocity offsets of the peaks from the line center are approximated by (e.g., \citealp{hansen2006,verhamme2008}):
\begin{equation}\label{eq:vOffStatic}
\Delta v \; \sim \; \pm \: 190 \: \left[\mathrm{km \; s}^{-1}\right] \: \left(\frac{b}{80 \: \mathrm{km \: s}^{-1}}\right)^{1/3} \left(\frac{N_{\mathrm{HI}}}{10^{19} \: \mathrm{cm}^{-2}}\right)^{1/3} \: .
\end{equation}
Imparting a bulk velocity to the slab towards the observer (away from the \lya\ source) results in a growing asymmetry in the flux of the two peaks as the \lya\ photons upon the first interaction are redshifted in the frame of the hydrogen atoms. As a result, bluer photons shift into resonance, which diminishes the bluer peak relative to the redder peak. When the slab has a bulk velocity, the velocity offsets of the emission peaks are no longer well approximated by the relation in Equation \ref{eq:vOffStatic}, but the basic behavior remains the same (i.e., the total separation between the peaks increases with increasing $b$ and \nh). This scenario qualitatively describes escape mode 1 for photons escaping the expanding shell on the hemisphere approaching the observer at \vexp, and the basic trends described above can be seen in the panels of Figure \ref{fig:ExModelSpec}. 

As mentioned, escape modes 1 and 2 together result in two emission peaks redward of the \lya\ line center. While several models in Figure \ref{fig:ExModelSpec} show this morphology, the majority only show a single redward peak. As noted by \citet{verhamme2006}, decreasing the ratio $V_{\mathrm{exp}} / b$ results in these two red peaks becoming increasingly superposed, eventually resulting in a single blended redward emission peak. Decreasing this ratio also results in a decrease of the ratio of the flux redward of the line center to the blueward flux. This is in part due to the move closer to the static case with decreasing \vexp. 

	\begin{figure*}[t]
	\begin{center}
	\begin{tabular}{c}
	\includegraphics[width=17cm]{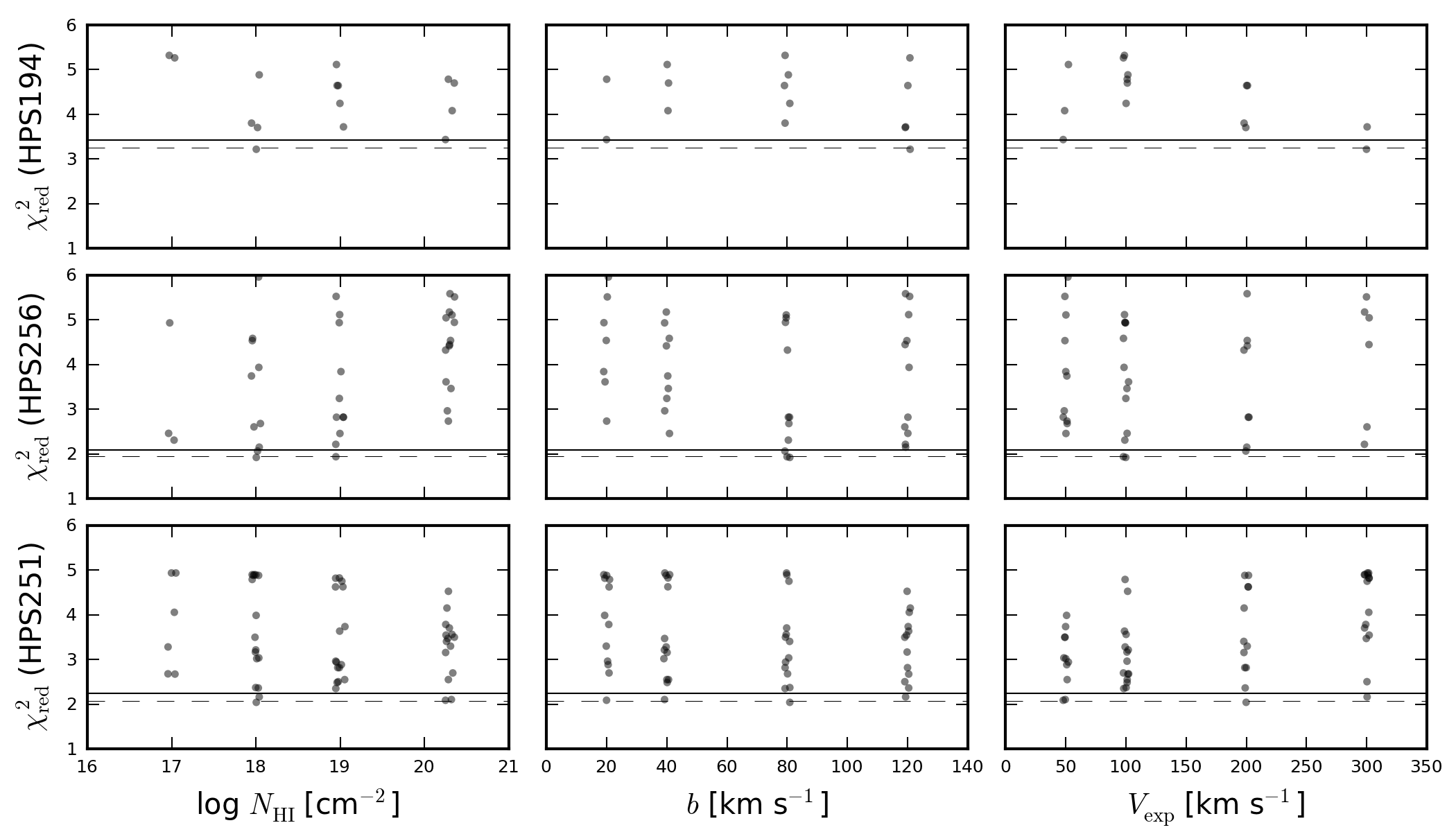}
	\end{tabular}
	\end{center}
	\caption[example] 
	{ \label{fig:Chi2Comparison} 
	The results of the statistical comparison of the \lya\ spectra of the three LAEs with the grid of model \lya\ line profiles in the expanding shell geometry. The reduced \chisq\ statistic ($\chi_{\mathrm{red}}^{2}$) is shown as a function of each of the three model parameters. In each panel, the horizontal dashed and solid lines represent the 68\% and 99\% confidence limits shown in the reduced units, respectively. To emphasize the minimum $\chi_{\mathrm{red}}^{2}$ model, the vertical axis of each panel is set such that models with $\chi_{\mathrm{red}}^{2} > 6$ are not shown. Each data point has been scrambled by a small, random amount along the horizontal axis and displayed with some degree of transparency to avoid confusion between similar-valued points.}
	\end{figure*} 

Changes in \nh\ result in complex variations of the line profile. In general, as roughly described by Equation \ref{eq:vOffStatic}, increasing \nh\ results in an increased separation between the peaks. Additionally, the FWHM and asymmetry of each peak increases with increasing \nh\ due to the photons needing to scatter further out into the line wings in order to escape the optically thick column, especially for the redward emission as the mode 2 backscattering effect is increased. This is also the cause of the decreasing ratio of the flux redward of the line center to the blueward flux for a given \vexp. As mentioned above, the velocity offset of the red peak that results from escape mode 2 (i.e., backscattered photons) typically traces $\sim2 V_{\mathrm{exp}}$. At first glance, this feature of the emergent \lya\ line profile seems to provide a tight constraint on the outflow velocity of the system. However, we've mentioned that this peak can become blended with the redward peak resulting from escape mode 1 for decreasing $V_{\mathrm{exp}} / b$. To further complicate the matter, \citet{verhamme2006} note that the redward peak from escape mode 1 decreases in strength relative to the escape mode 2 peak with increasing \nh. This effect is clearly illustrated in Figure 17 of \citet{verhamme2008}, which shows how the velocity offset of the \textit{dominant} redward \lya\ peak traces a varying multiplicative factor of \vexp\ with various values of \nh. For $N_{\mathrm{HI}} \gtrsim 10^{20}$ cm$^{-2}$, the most dominant of the two redward peaks results from the backscattered escape mode 2 photons and traces $\sim2 V_{\mathrm{exp}}$. For smaller \nh, the dominant redward peak results from escape mode 1 photons and its velocity offset traces $<2 V_{\mathrm{exp}}$. This explains why little change in the velocity offset of the dominant red peak in the right panel of Figure \ref{fig:ExModelSpec} is seen with changing \vexp. Since the models we show are for \nh\ $= 10^{19}$ cm$^{-2}$, the dominant peak results from escape mode 1 photons whose velocity offset from the line center is primarily determined by changes in \nh\ and $b$ (cf., Equation \ref{eq:vOffStatic}). \nh\ is particularly dominant because, in addition to shifting the emission components in velocity, it has the ability to strongly influence the FWHM and asymmetry of the emission components. Additionally, its dynamic range is much larger than that for $b$. 

As can be seen by the above discussion and the example model spectra shown in Figure \ref{fig:ExModelSpec}, the emergent \lya\ line profile changes in non-trivial fashion for variations in each individual parameter. In addition, simple qualitative interpretations of observed \lya\ line profiles under the expanding shell model can be hindered by the degeneracies of the modeling parameters on all observables (e.g., one needs to constrain \nh\ before using the velocity offset of the dominant redward peak directly as a tracer of \vexp). While such constraints are not currently available for our galaxies, our observed \lya\ spectra are of sufficient resolution to identify individual emission components and their shapes (i.e., asymmetries, widths, etc.). By conducting a statistical comparison of our observed \lya\ spectra with the spectra of the expanding shell model grid, we aim to constrain the physical parameters of the gaseous component of the LAEs through which the \lya\ photons scatter. 

	\begin{figure*}[t]
	\begin{center}
	\begin{tabular}{c}
	\includegraphics[width=18cm]{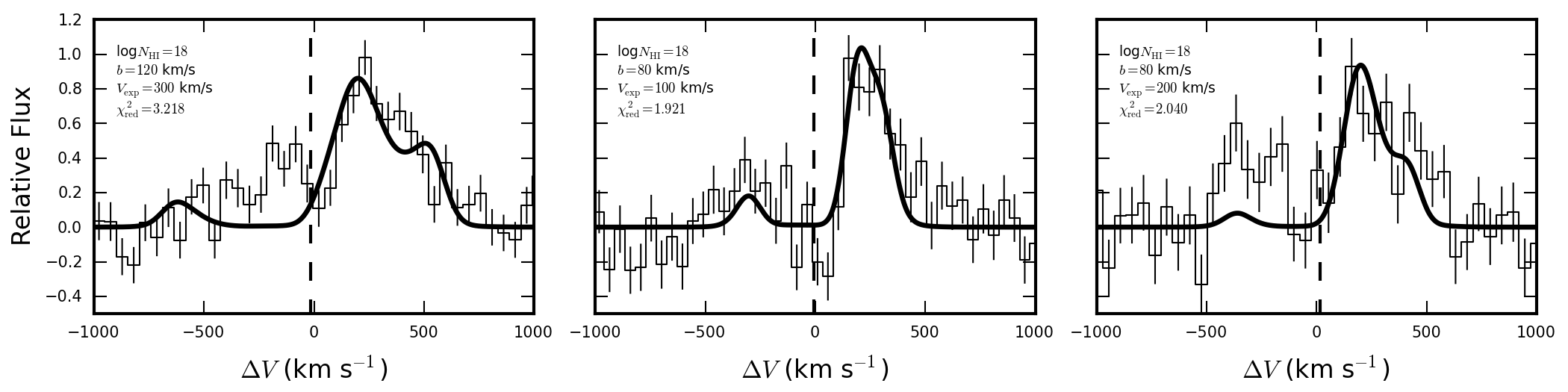}
	\end{tabular}
	\end{center}
	\caption[example] 
	{ \label{fig:ModelComparison} 
	The observed \lya\ spectra of the three LAEs plotted with their respective best-fitting expanding shell model \lya\ spectra. The magnitude of the velocity shift within the total velocity zero-point uncertainty that yields the minimum \chisq\ is visualized in these panels by the shifted vertical dashed line relative to $\Delta v = 0$ km s$^{-1}$. The best-fitting models and their respective statistics are listed in Table \ref{table:Chi2Summary}.} 
	\end{figure*} 

\subsection{Statistical Comparison with Observed \lya}\label{subsec:ModelCompare}
We perform a statistical comparison between the observed \lya\ emission line profiles and the expanding shell \lya\ radiative transfer models by calculating the \chisq\ statistic for each model on the grid for each galaxy. After being degraded to the instrumental resolution, each model in the grid is interpolated via cubic spline to the velocity bin centers of each galaxy's observed \lya\ spectrum. The observed spectra are corrected for the background level $C$ measured in $\S$\ref{subsubsec:LyAQuantitative} (this is important only for HPS251, which includes faint continuum from the nearby extended low-redshift galaxy that was within the fiber; see Figures \ref{fig:PointingThumbs} and \ref{fig:VSpec}). Finally, we multiply the model by a scaling factor $A$, which is determined by minimization of the squared residuals between the model and the data. We calculate the \chisq\ statistic as follows:
\begin{equation}\label{eq:chisquare}
\chi^{2} \; = \; \sum^{N}_{i = 1} \: \left[\frac{F_{i} \; - \; y\:(N_{\mathrm{HI}},\: b,\: V_{\mathrm{exp}},\: A;\: \Delta v_{i})}{\sigma_{i}}\right]^{2} \: ,
\end{equation}
where $F_{i}$ and $\sigma_{i}$ are the observed relative flux and its associated statistical uncertainty, respectively, each at the $i^{\mathrm{th}}$ velocity bin $\Delta v_{i}$. Here, the model spectrum with the parameters \nh, $b$, \vexp, and $A$, is denoted as $y$. Recall that there is an uncertainty on the velocity zero-point of the observed \lya\ spectrum that results from the RMS of the Mitchell Spectrograph and NIRSPEC wavelength solutions as well as the systematic uncertainty associated with the $z_{\mathrm{sys}}$ measurement (see $\S$\ref{subsec:VPobs} and $\S$\ref{subsubsec:NIRSPEC}). To reflect this uncertainty in our analysis, we have allowed the observed \lya\ spectra to shift relative to the velocity zero-point within this total uncertainty until the minimum \chisq\ is found for each model on the grid. The total uncertainty in the velocity zero-point is $\pm19$, $\pm20$, and $\pm21$ km s$^{-1}$ for HPS194, HPS256, and HPS251, respectively.

	\begin{deluxetable}{ccccccc}
	\tabletypesize{\scriptsize}
	\tablecaption{Summary of Best-Fit Expanding Shell Models\label{table:Chi2Summary}}
	\tablehead{ \colhead{Object} & \colhead{$N-M$} & \colhead{$Q$} & \colhead{$\log{N_{\mathrm{HI}}}$} & \colhead{$b$} & \colhead{$V_{\mathrm{exp}}$} & \colhead{$\chi_{\mathrm{red}}^{2}$}\\ &	&	&  & \colhead{ (km s$^{-1}$)} & \colhead{ (km s$^{-1}$)} & \\}
	\startdata
	HPS194 & 33 & $4.1\times10^{-10}$ & 18 & 120 & 300 & 3.218 \\[0.7ex]
	HPS256 & 39 & $7.8\times10^{-4}$ & 18 & 80 & 100 & 1.921 \\[0.7ex]
	HPS251 & 34 & $3.2\times10^{-4}$ & 18 & 80 & 200 & 2.040 \\[0.7ex]
	\enddata
	\tablecomments{The models listed are the best-fit at the 68\% confidence level, with the exception of HPS256. For HPS256, a model with the same $b$ and \vexp\ (but with $N_{\mathrm{HI}} = 10^{19}$ cm$^{-2}$) is the next best-fit. Since these two models differ by $\Delta\chi^{2} = 0.61$., they both lie within the 68\% confidence limit.}
	\end{deluxetable}

For a common comparison of the calculated \chisq\ values between the three galaxies, we calculate the reduced \chisq\ statistic, which is given by $\chi_{\mathrm{red}}^{2} = \chi^{2} / (N - M)$, where $N$ is the number of velocity bins in each spectrum and $M$ is the number of degrees of freedom (here, $M=5$). Since we only included data in the range $-1000 < \Delta v$ (km s$^{-1}$) $< 1000$ in our calculation of \chisq\ (i.e., we ignore the second blueward peak in HPS256 and HPS251), $N$ is 38, 44, and 39 for HPS194, HPS256, and HPS251, respectively. 

In Figure \ref{fig:Chi2Comparison}, we show the results of the statistical comparison of the $R\approx2500$ data with the expanding shell model grid by plotting $\chi_{\mathrm{red}}^{2}$ as a function of each of the three physical model parameters for each galaxy. We include horizontal lines that represent the $\Delta\chi^{2}$ above the minimum \chisq\ for each galaxy in the reduced units that corresponds to the 68\% and 99\% confidence limits \citep{press1992}. As can be seen from Figure \ref{fig:Chi2Comparison}, the data appear to be able to constrain the models relatively well at the 68\% confidence level. For HPS194 and HPS251, a single best-fit model lies below the 68\% confidence limit. For HPS256, two models lie below the 68\% confidence limit, although these two models occupy nearly same position in the three dimensional parameter space and differ only by one order of magnitude in \nh. At the 99\% confidence level, HPS194 still only has a single best-fitting model. HPS256 and HPS251, however, have three and four models lying below the 99\% confidence limit, respectively, and are thus less well constrained. In particular, the lower $S/N$ of the HPS251 \lya\ spectra resulted in the four models within the 99\% confidence limit spanning the entire parameter space in both $b$ and \vexp.

The best-fit models for each galaxy are summarized in Table \ref{table:Chi2Summary} and are plotted with the $R\approx2500$ \lya\ data in Figure \ref{fig:ModelComparison}. From this figure, we can qualitatively state that the emission redward of the \lya\ line center is relatively well represented by the expanding shell model spectra for each galaxy except for the extended red wings for HPS256 and HPS251. The emission blueward of the \lya\ line center, however, is very poorly fit by the best-fitting model for each galaxy except for HPS256. For HPS194 and HPS251, the models do not reproduce the proper velocity offset or red-to-blue peak flux ratio. To estimate the probability of the observed residuals being due to statistical fluctuations in the data, we have calculated the probability $Q$, which is given by the integral of the \chisq\ probability density function for $N-M$ degrees of freedom, integrated from the best-fit \chisq\ value for each galaxy to infinity \citep{press1992}. We obtain probabilities of $4.1\times10^{-10}$, $7.8\times10^{-4}$, and $3.2\times10^{-4}$ for HPS194, HPS256, and HPS251, respectively. This suggests that even the best-fitting expanding shell model \lya\ spectra are relatively poor representations of the observed \lya\ spectra for these galaxies.

\section{DISCUSSION}\label{sec:Discussion}
\subsection{Possible Limitations of Our Models}\label{subsec:Limitations}
In $\S$\ref{subsec:ModelGrid}, we discussed the differences between our homogeneous expanding shell models and those of \citet{verhamme2006} and \citet{schaerer2011}. The differences are the simplification of the input spectrum and the lack of dust. We discuss each of these below as they pertain to the statistical comparison presented above. 

We have already stated that the lack of continuum is negligible in our models since our objects have such high \ewlya. However, all three galaxies have a measurable value of $\sigma_{\mathrm{H}\alpha}$ from the spectrally resolved \ha\ emission (Song et al. 2013, in preparation). Assuming the \lya\ photons are originally emitted from the same gas that is emitting the detected \ha, these measurements imply an intrinsic \lya\ FWHM of $144\pm22$, $170\pm22$, and $104\pm14$ km s$^{-1}$ for HPS194, HS256, and HPS251, respectively. A higher intrinsic \lya\ line width (as opposed to the monochromatic line we model) results in more photons beginning the resonant scattering process further from the core of the scattering cross-section. In non-static media, however, the frequency of the resonant core is shifted in the frame of the scattering gas. Thus, for the expanding shells we model, a non-monochromatic intrinsic \lya\ line increases the direct escape probability of photons originally emitted redward of the \lya\ line center while decreasing the corresponding probability for blueward emitted photons. The result is a slightly broader \lya\ spectrum with similar velocity offsets and a slightly increased value of $F_{red} / F_{blue}$. A broader emergent \lya\ spectrum would aid in fitting the extended red wings of the redward emission component for HPS256 and HPS251. However, this improvement may be negatively offset because, as seen in Figure \ref{fig:ModelComparison}, the flux ratio $F_{red} / F_{blue}$ is consistently too large in the models. In addition, a non-monochromatic intrinsic \lya\ emission line would not significantly change the model emission component velocity offsets. Since the velocity offset of the HPS194 and HPS251 blueward emission component is already reproduced incorrectly by \lya\ photon transfer through the expanding shell, including a frequency distribution rather than monochromatic \lya\ photons would likely result in only negligible improvements for those objects.

As seen in $\S$\ref{subsec:ModelGrid}, each LAE contains potentially significant amounts of dust (although the large uncertainties on $E(B-V)$ are also consistent with very little or no dust; \citealp{blanc2011}). \citet{verhamme2006} discuss how the various features of the \lya\ line profile that arise from the various escape modes (see $\S$\ref{subsec:GridTrends}) change with increasing dust optical depth $\tau_{a}$. In general, the photons that encounter more scatterings have a higher probability of being absorbed by dust due to their longer effective path length through the system. Thus, the backscattered photons of escape mode 2, particularly the extended red tail resulting from multiple backscattering events, are especially affected by dust which results in a more narrow (or ``sharpened'') line profile (also, see \citealp{laursen2009}). Additionally, the photons emerging blueward of the line center (i.e., the bluer photons resulting from escape mode 1 scatterings through the hemisphere of the shell expanding towards the observer) are also especially susceptible to dust absorption. This is because the bluer photons are redshifted closer to the resonant frequency as seen by the hydrogen atoms in the shell and undergo a larger number of core scatterings before emerging blueward enough for escape. These effects can be seen in Figure 17 of \citet{verhamme2006}, Figures 5 and 6 of \citet{schaerer2011}, and Figure 16 of \citet{duval2013}. As mentioned above, the expanding shell models within the searched parameter space do not reproduce the observed values of $F_{red} / F_{blue}$, with the peak blueward of the line center ubiquitously having too little flux in the best-fit models. Since the blueward peak is one of the features that is most easily extinguished by dust, including dust in our models cannot improve the best-fitting values of \chisq\ since the little flux in the models blueward of the line center would only be further diminished.

As is often the case at $2\lesssim z \lesssim 3$, we have ignored the effect of absorption and scattering in the IGM.  While its effect is mostly to further diminish the \lya\ emission that is blueward of the \lya\ line center (e.g., \citealp{laursen2011}), it is possible for gravitational inflow of IGM material in the vicinity of the galaxy's dark matter halo to also diminish the redward emission \citep{dijkstra2006b}. However, since the mean optical depth of the IGM is $\lesssim0.2$ at the redshifts of the galaxies included in this study \citep{becker2013}, we do not consider it as a contributing factor in the poor fits to the data. 

Finally, we reiterate that we have modeled \lya\ radiative transfer through expanding shells with a \textit{homogeneous} distribution of neutral gas. \citet{duval2013} have modeled \lya\ radiative transfer through expanding shells with varying degrees of density inhomogeneity, from the homogeneous case to a scenario where all of the neutral gas is contained in clumps (i.e., the volume in the shell between the clumps is transparent to \lya\ photons). In general, they find that the overall shape of the emergent \lya\ line profile for increasing degrees of inhomogeneity is basically unchanged and follows the same trends as seen in the homogeneous case (see $\S$\ref{subsec:GridTrends} and \citealp{verhamme2006}). However, one marked change for parameter combinations resulting in double-peaked spectra (i.e., models with low $V_{\mathrm{exp}} / b$) is the reduction in $F_{red} / F_{blue}$ with increasing inhomogeneity. As seen in Figure 13b of \citet{duval2013} for an expanding shell with $N_{\mathrm{HI}} = 2\times10^{20}$ cm$^{-2}$, $b = 40$ km s$^{-1}$, and $V_{\mathrm{exp}} = 100$ km s$^{-1}$, $F_{red} / F_{blue}$ reduces from $\sim11$ in the homogeneous case to $\sim5$ in the clumpy case. Since our homogeneous models consistently produce too large $F_{red} / F_{blue}$, this effect would certainly improve the overall fit of the expanding shell model spectra to our observed \lya\ spectra. 

While a highly inhomogeneous, multi-phase ISM is not fully consistent with LAE observations and may be astrophysically unrealistic (e.g., \citealp{scarlata2009,duval2013,laursen2013}), several recent studies have shown that some degree of inhomogeneity is present for the average LAE (e.g., \citealp{blanc2011,nakajima2012}). These studies derive the parameter $q = \tau_{\mathrm{Ly}\alpha} / \tau_{1216}$, where $\tau_{\mathrm{Ly}\alpha}$ is the optical depth of \lya\ photons and $\tau_{1216}$ is the optical depth of continuum photons redward of the \lya\ transition. Values of $q \ll 1$ indicate that \lya\ sees very little dust extinction, as would be the case for a reduced number of scatterings in an ISM that has either favorable kinematic properties (e.g., the gas experiences a high velocity bulk motion, such as a powerful outflow), a highly inhomogeneous and clumpy distribution of neutral gas and dust (e.g., \citealp{neufeld1991,hansen2006}), or both. Values of $q \gg 1$ indicate that \lya\ photons suffer a large number of scatterings and are thus strongly attenuated by dust, which is expected for a homogeneous and/or static ISM. For a sample of $\sim900$ LAEs at $z=2.2$, \citet{nakajima2012} calculated $\langle q \rangle = 0.7 \pm 0.1$, while \citet{blanc2011} measure a median $q$ of $0.99\pm0.44$ for LAEs in the HPS sample. Both of these measurements indicate that on average for a similar LAE to those we observe, \lya\ photons are neither preferentially attenuated nor are preferentially escaping due to the configuration of the ISM. Although the effects of the ISM's distribution and kinematics both factor into the measured value of $q$, $q\approx1$ indicates that some degree of ISM inhomogeneity cannot be ruled out.

In summary, the limitations of our \lya\ radiative transfer modeling may be a factor in explaining some of the discrepancies between the model \lya\ spectra and our observations. For HPS256, the expanding shell model properly reproduces the velocity offsets of the \lya\ emission components. Including a distribution of frequencies for the input \lya\ photons (which would create a broader emergent \lya\ spectrum, better fitting the extended red wing of the red emission component) and modeling an inhomogeneous gas distribution within the expanding shell (which would decrease $F_{red} / F_{blue}$) would likely result in a model spectrum that closely matches our observations. However, such modeling advancements will not improve the fit when the velocity offsets of the \lya\ emission components are not properly reproduced in the first place (e.g., the blue peak of HPS194 and HPS251).

\subsection{Expanding Shells: the Right Model?}\label{subsec:ShellsCorrect}
The expanding spherical shell model describing gaseous geometry and kinematics is appealing for star-forming galaxies due to its simplicity and physical motivation. This model has recently received a great deal of attention in the literature from an observational standpoint for interpreting \lya\ line profiles of LAEs and LBGs and their velocity offsets. \citet{verhamme2008} had success in fitting the \lya\ line profiles of a sample of 11 LBGs (8 have \ewlya\ $> 20$ \AA) observed at $R\approx2000$ with expanding shell synthetic spectra. Their spectra were typically constrained by systemic redshift measurements of non-resonant emission lines in the low resolution FORS Deep Field spectra of \citet{noll2004}. They use the physical parameters derived from those fits to explain several observed correlations between various properties of the sample. Of their 11 galaxies, 7 display a \lya\ line with a single asymmetric peak. Their spectra are typically deep enough to detect the continuum, so secondary peaks are not missed due to $S/N$ limitations. The remaining 4 galaxies show multiple-peaked morphologies similar to the \lya\ spectra of the LAEs presented here. Unlike for the single-peaked asymmetric profiles, \citet{verhamme2008} have difficulty fitting the double-peaked \lya\ profiles with the expanding shell model and typically require quasi-static gas kinematics, much larger intrinsic \lya\ line widths than are physical for non-AGN star-forming galaxies, or a large adjustment of the velocity zero-point (of $\gtrsim200$ km s$^{-1}$) in order to obtain a good fit. 

Observing LBGs with existing \lya\ data at an average spectral resolution of $\sim370$ km s$^{-1}$ FWHM, \citet{kulas2012} specifically targeted objects displaying multiple-peaked \lya\ spectra and followed-up on a sample of 18 objects with NIR observations of optical nebular emission lines to constrain the sample's $z_{\mathrm{sys}}$. They observe various \lya\ line profile morphologies and qualitatively compare composite spectra grouped by morphology with a coarse grid of synthetic spectra from the same expanding shell \lya\ radiative transfer code used here. In general, they found that most features of the various multiple-peaked spectra could not be reproduced by the models. The best match to the expanding shell model is their ``Group 1'' profiles, which are qualitatively similar to the \lya\ profiles in this work (see Figure \ref{fig:K12G1Comparison}). A high column density expanding shell model with \nh\ = $2\times10^{20}$ cm$^{-2}$, $b = 40$ km s$^{-1}$, and \vexp\ = 100 km s$^{-1}$ reproduces the large velocity offsets for their composite ``Group 1'' LBG profile as well as the asymmetries. But, like we observe in Figure \ref{fig:ModelComparison}, the expanding shell model under-predicts the strength of their ``Group 1'' composite's blue emission component. \citet{kulas2012} also measure the widths and velocity offsets of IS absorption features. Assuming that the absorbing material is part of the same outflowing shell through which \lya\ photons scatter, the former measure traces $b$ while the latter is equivalent to \vexp. Indeed, they find qualitatively that the best-fitting model spectrum has \vexp\ = 100 km s$^{-1}$, which is comparable to the 90 km s$^{-1}$ blueshift of the ``Group 1'' IS absorption lines. However, their best-fit value of $b$ is lower than that estimated from the observed IS absorption line widths for those objects by a factor of $\sim7$.  

In addition to the aforementioned discrepancies between predicted \lya\ line profiles from the expanding shell model and observed double-peaked \lya\ profiles, the deep spectroscopy of faint LAEs from \citet{rauch2008} has allowed for the study of spatial \lya\ surface brightness (SB) profiles. The \citet{rauch2008} sample is comprised of several single-peaked, asymmetric \lya\ spectral line profiles as well as some that are double-peaked. \citet{barnes2010} have predicted the SB profiles using radiative transfer models with various spatial and velocity configurations of the gas, including expanding shells. They find that \lya\ radiative transfer through an expanding shell typically results in a flat spatial SB profile, which is at odds with the peaky composite profile of the faint LAEs in the \citet{rauch2008} sample. While the faint LAEs of \citet{rauch2008} may be in a different class of objects than the bright LAEs we observe, \citet{rauch2011} have performed a similar comparison to \citet{barnes2010} with the spatial \lya\ SB profiles of single, bright LBGs having both single and double-peaked spectral line profile morphologies. In both cases, \citet{rauch2011} determine that the flat spatial SB profile resulting from an expanding shell is inconsistent with the peaky \lya\ SB profiles with extended wings that are observed. They conclude that better representations of the observed \lya\ SB profiles for bright LAEs are found by modeling a point source of ionizing radiation within an optically thick, slowly expanding halo of neutral gas rather than a shell. 

In this work, we have provided examples of LAEs whose double-peaked \lya\ spectral line profiles are not well-reproduced by radiative transfer through \textit{homogeneous} expanding shells. The largest discrepancies are the flux ratio $F_{red} / F_{blue}$ and the velocity offset of the blue component. Extrapolating from Figure \ref{fig:ExModelSpec}, one can decrease $F_{red} / F_{blue}$ \textit{and} reduce the blue component velocity offset qualitatively in better agreement with the observed \lya\ line profiles of these objects by decreasing the shell's \vexp\ well below 50 km s$^{-1}$, which is the smallest value in our grid. For this near-static case, the total velocity separation between the two emission components straddling the systemic \lya\ line center are to first-order given by twice the value calculated using Equation \ref{eq:vOffStatic}. From Equation \ref{eq:vOffStatic}, it can be seen that reproducing the observed $\Delta v_{\mathrm{tot}}$ for essentially any value of $b$ requires $N_{\mathrm{HI}} \sim 10^{-19}$ cm$^{-2}$, which is too low of a column density to simultaneously reproduce the large FWHM of each peak \textit{and} the extended red wings in the observed \lya\ data (see Figures \ref{fig:ExModelSpec} and \ref{fig:ModelComparison}). Additionally, the \ha\ derived star formation rates (SFR; uncorrected for dust; see Table \ref{table:GalData}) for these three galaxies are large for a typical LAE. Combined with their small sizes ($\lesssim1.6$ kpc; see $\S$\ref{subsec:Imaging}) and dynamical masses (see \citealp{rhoads2013}), the resulting SFR surface density $\Sigma_{\mathrm{SFR}}$ of these galaxies indicates that they should be driving strong outflows of $>50$ km s$^{-1}$ if \vexp\ is comparable to the escape velocity \citep{heckman2002,newman2012}. This strongly disfavors any quasi-static shell model. 

Finally, to this point, we have ignored the second blue emission peak in the spectrum of HPS256 and HPS251 and assumed that these two galaxies represent typical double-peaked LAE. If this assumption is invalid and the second blue peak is \lya\ emission from the same system, the expanding shell geometry can most likely be immediately ruled out as the geometric and kinematic configuration of the neutral gas distribution for these objects. Triple-peaked \lya\ emission profiles can be produced by the expanding or infalling\footnote[14]{The synthetic \lya\ spectra produced by the shell geometry expanding at \vexp\ are also valid for a shell infalling with the same velocity, in which the \lya\ profile is reversed about $\Delta v = 0$ km s$^{-1}$ relative to the expanding case \citep{verhamme2006}.} shell geometry (cf., Figure \ref{fig:ExModelSpec}), but not with the observed relative strengths of the various components when considering their respective locations relative to the velocity zero-point.  

\subsection{Other Models}\label{subsec:OtherModels}
The aforementioned discrepancies between the expanding shell model and the observations should lead us towards considering different, or more complex gas geometries and velocity fields. These differences should include deviations away from spherical symmetry, unity covering factors, and strictly outflowing gas. Works such as \citet{christensen2012} and \citet{noterdaeme2012} incorporate multi-phase media in their gas geometries to closely reproduce multiple-peaked \lya\ line profiles. The former work is actually an expanding shell (whose parameters are within the space covered by our grid: \nh\ $\approx 10^{18}$ cm$^{-2}$, $b$ $\approx 90$ km s$^{-1}$, \vexp $\approx 50 - 100$ km s$^{-1}$) that includes dense neutral clouds distributed within a more ionized and less dense ``intercloud medium'' inside the shell's cavity. The latter work incorporates an overall inflowing velocity field with starburst driven bipolar outflowing gaseous ``jets'' to account for the strong blueward \lya\ emission they observe for a double-peaked line profile. A similar model was adopted by \citet{adams2009} to constrain the spatially resolved 2-D \lya\ emission from a $z=3.4$ radio galaxy, where ionized cones aligned with the galaxy's radio axis are embedded in an infalling neutral halo. 

Current state-of-the-art modeling efforts are post-processing galaxy models drawn from hydrodynamic and cosmological galaxy formation simulations with \lya\ radiative transfer codes (e.g., \citealp{zheng2010,kollmeier2010,barnes2011,verhamme2012,yajima2012}). For example, \citet{barnes2011} find that a typical halo contains a mixture of inflowing and outflowing gas and that the relative contribution of each along the line of sight determines the relative strength of the \lya\ emission blueward (for inflow) and redward (for outflow) of the velocity zero-point. As a result, the \lya\ line profile can vary as a function of viewing angle for the same halo. Also, \citet{verhamme2012} post-processed high resolution hydrodynamical simulations that follow the formation and evolution of isolated disk galaxies with a \lya\ radiative transfer code. Their simulations resolve the small-scale structure of the ISM (including the thick, fragmented clouds in which stars form and the \lya\ photons originate), which they find to be extremely important in determining the galaxy's \lya\ properties. The clumpy disk galaxy they model harbors an axially asymmetric, large-scale outflowing velocity field that is mostly perpendicular to the disk. The asymmetric outflow aids in the escape of \lya\ photons and results in galaxy inclination strongly affecting the observed \lya\ emission in terms of the escape fraction and \ewlya, the \lya\ SB distribution, and the shape of the \lya\ line profile. For edge-on orientations, they expect to observe more symmetric double-peaked \lya\ spectra as a result of little-to-no outflow along the line of sight and the higher optical depth to the star-forming regions through the disk. For increasingly face-on orientations, increasing \ewlya\ and asymmetry between the two emission peaks should develop as the outflowing gas perpendicular to the disk gains a larger velocity component along the line of sight. While typical star-forming galaxies at $z\sim2.4$ are likely not disks in the classical sense (e.g., \citealp{law2012a}) and probably do not have bipolar outflows (e.g., \citealp{law2012b}), the more realistic treatment of galactic systems by \citet{verhamme2012} has shown the importance of considering \lya\ radiative transfer on small-scales within the ISM. Expanding shells modeling large-scale galactic outflows do not capture such physics, which may be an additional cause for the discrepancies we observe between the expanding shell model predictions and the spectrally resolved \lya\ line profiles of the three LAEs.

	\begin{figure*}[t]
	\begin{center}
	\begin{tabular}{c}
	\includegraphics[width=17cm]{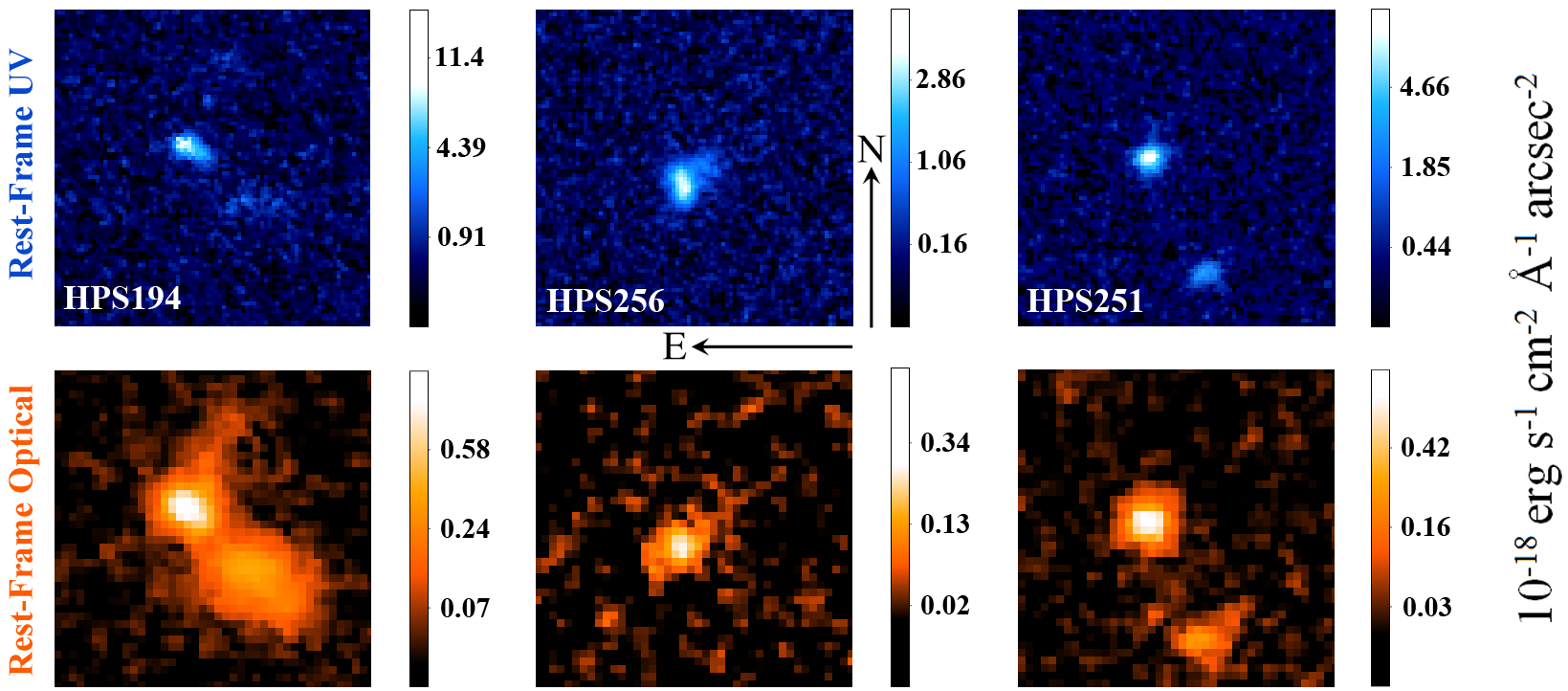}
	\end{tabular}
	\end{center}
	\caption[example] 
	{ \label{fig:COSMOSThumbs} 
	CANDELS \textit{HST} images \citep{koekemoer2011} of the three LAEs, which show the stellar continuum morphologies of each galaxy. The top row shows ACS/WFC F606W images (which probe the rest-frame UV at $\sim$1800 \AA) and the bottom row shows WFC3/IR F160W images (rest-frame optical at $\sim$4700 \AA). Each panel is $2.5\arcsec\times2.5\arcsec$ ($21\times21$ kpc at $z\sim2.4$) in size, and their positions are indicated in the wider field Subaru $V$ band images shown in Figure \ref{fig:PointingThumbs} by the thick black boxes.}
	\end{figure*} 

\subsection{Does $\Delta v_{\mathrm{Ly}\alpha}$ Indicate the Magnitude of $V_{\mathrm{exp}}$?}\label{subsec:Correlation?}
Many studies have used the results of \lya\ radiative transfer through expanding shells to help explain the observed velocity offsets of the \lya\ emission line from $z_{\mathrm{sys}}$ for star-forming galaxies and further to constrain the velocity of their large-scale outflows (e.g., \citealp{mclinden2011,finkelstein2011,yang2011,hashimoto2012,guaita2013}; Song et al. 2013, in preparation). The analysis in $\S$\ref{subsec:ModelCompare} of the spectrally resolved, multiple-peaked \lya\ emission of the three LAEs we present shows that it is difficult to constrain some of the physical parameters of each galaxy's outflowing gas. This is especially true since even the best-fitting expanding shell radiative transfer models show major discrepancies when compared to the resolved \lya\ spectra. Given these discrepancies, it is of note that \nh\ and $b$ appear to be much better constrained than \vexp. For each galaxy, the best-fitting model has $N_{\mathrm{HI}} = 10^{18}$ cm$^{-2}$ and a high value of $b$ as a result of their strikingly similar values of $\Delta v_{0,red}$ and $\mathrm{FWHM}_{red}$. In addition, as can be seen in Figure \ref{fig:ExModelSpec}, there is little change in the position of the dominant redward \lya\ peak over the range $50 < V_{\mathrm{exp}}$ (km s$^{-1}$) $< 200$ for an expanding shell model with low neutral column density (i.e., $N_{\mathrm{HI}} \lesssim 10^{20}$ cm$^{-2}$; see $\S$\ref{subsec:GridTrends}). Especially since this strong redward peak dominates the observed \lya\ line profile at low spectral resolution and/or $S/N$, it appears that \vexp\ has the least influence on $\Delta v_{\mathrm{Ly}\alpha}$ of the three physical parameters that we model for a galaxy with low neutral gas column density.

Several of the aforementioned studies (\citealp{mclinden2011,finkelstein2011,guaita2013}; Song et al. 2013, in preparation) do in fact utilize a lower spectral resolution that is similar to the HPS when observing \lya\ such that objects with line profiles similar to those we present here are unresolved (see Figure \ref{fig:VSpec}). In cases for objects with multiple-peaked \lya\ line profiles with $\Delta v_{\mathrm{tot}}$ less than the instrumental resolution, much of the information that is encoded in the multiple-peaked profile is lost and the only measurable quantity from the line profile becomes $z_{\mathrm{Ly}\alpha}$ (or $\Delta v_{\mathrm{Ly}\alpha}$ when a measure of $z_{\mathrm{sys}}$ exists). We have also previously noted the instrumental effect of observing an intrinsically asymmetric emission line at low resolution in which the peak flux can can be biased in the direction of the extended emission tail (see $\S$\ref{subsubsec:LyAQuantitative}). As a result of the loss of information, the measured $\Delta v_{\mathrm{Ly}\alpha}$ for an unresolved line becomes a convoluted function of many parameters, including the line of sight velocity field, column density, and the temperature of the neutral gas in addition to parameters that were not explored in our models, such as dust content. Even when ignoring the possible effects of the IGM on the \lya\ line profile (e.g., \citealp{dijkstra2006b,laursen2011}), the matter is further complicated when considering \lya\ radiative transfer through more realistic models of the gas distribution and velocity fields for galaxies (see the previous subsection). These models, which do not have spherically symmetric gas distributions, indicate that viewing angle should also be an important parameter in determining the degree of flux asymmetry between the multiple peaks, which can affect the value of the measured $\Delta v_{\mathrm{Ly}\alpha}$. 

As a result of all of these factors, attempts to equate the velocity offsets from systemic that are observed in \lya\ spectra to physical outflow speeds need to be treated with extreme caution. We also stress the importance of observing \lya\ with high spectral resolution in order to extract the maximum amount of information from the line profile. We do point out, however, that while the \textit{magnitude} of the inflow or outflow speed is not yet clearly a measurable quantity from the \lya\ line profile even when spectrally resolved, the detection and sign of a non-zero $\Delta v_{\mathrm{Ly}\alpha}$ does at least indicate presence of an outflow (neglecting IGM absorption) or potentially inflow (for blueshifted \lya\ emission) along the line of sight. 

\subsection{Clues from Spatially Resolved Data}\label{subsec:Imaging}
In Figure \ref{fig:COSMOSThumbs}, we show CANDELS \textit{HST} ACS/WFC F606W and WFC3/IR F160W images of the three LAEs, which probe the rest-frame UV ($\sim$1800 \AA; this wavelength traces recent star formation through the continuum emission of young, massive stars) and rest-frame optical ($\sim$4700 \AA), respectively. As expected from near-UV \textit{HST} morphological studies of LAEs at similar redshift (e.g., \citealp{bond2012,law2012b}), the galaxies are compact with half-light radii\footnote[15]{COSMOS ACS $I$-band Photometry Catalog, June 2008 Release} $r_{e} =$ 1.6, 1.1, and 0.7 kpc for HPS194, HPS256, and HPS251, respectively, at $\sim$2400 \AA\ \citep{leauthaud2007}. 

At the spatial resolution of the \textit{HST}, two of the galaxies (HPS194 and HPS251) are shown to have a companion continuum source at $\lesssim 1\arcsec$ distance:

\textit{HPS194:} This galaxy consists of a compact source to the north-west with a possible tidal tail reaching towards the north (this is visible as small clumps in the rest-frame UV image and is a very pronounced continuous feature in the rest-frame optical image). A more diffuse companion is located $\sim0.6\arcsec$ to the south-west in projection ($\sim5$ kpc if the sources are at the same redshift). As seen in Figure \ref{fig:PointingThumbs}, the NIRSPEC slit used by \citet{finkelstein2011} is aligned along the two sources. However, we were unable to detect \ha\ emission from the fainter source. As will be presented by Blanc et al. 2013 (in preparation), we have also obtained deeper NIR spectra of HPS194 with the FIRE instrument \citep{simcoe2013} at the 6.5 m Magellan Baade telescope. A preliminary reduction of the data yields no detection of the diffuse companion's rest-frame optical emission lines. Although we cannot confirm from the available data if the compact and more diffuse sources lie at the same redshift, the disturbed, asymmetric morphology and possible tidal feature emanating from the north-west compact source is suggestive of an ongoing merger.

\textit{HPS251:} This galaxy has a fainter, compact companion located $\sim1.0\arcsec$ ($\sim8$ kpc) to the south-south-west in projection. Like HPS194, the NIRSPEC slit used by Song et al. (2013, in preparation) is aligned along the two sources. Unfortunately, two of the positions of the \textit{ABBA} dither pattern positioned the dimmer source on the bottom edge of the slit, yielding insufficient $S/N$ in the final 2-D spectrum at the position of the companion. However, recent follow-up observations using FIRE have confirmed detections of \oiii\ $\lambda4959$ and \oiii\ $\lambda5007$ for both sources at the same redshift (these results will be presented by Blanc et al. 2013, in preparation). At the projected distance of $\sim8$ kpc, the two components of HPS251 are clearly interacting and will eventually merge.

While HPS256 does not have a companion within $1\arcsec$, it does have several nearby continuum sources within $\sim4\arcsec$ to the south-east in projection ($\sim33$ kpc if the sources are at the same redshift; see Figure \ref{fig:PointingThumbs}). However, these sources are outside of the Mitchell Spectrograph's fiber and the NIRSPEC slit used by \citet{finkelstein2011} was not aligned to include them. 

\citet{cooke2010} explored the \lya\ properties of close LBG pairs and found that all showed \lya\ in emission when the projected separation was $\lesssim 15$ kpc. Their work supports the picture of galaxy-galaxy interactions triggering star-formation and clearing gas and dust sufficiently for the \lya\ photons produced in the starburst to escape the galaxies with high \ewlya. \citet{cooke2010} also find that while the 1-D spectra of the close LBG pairs are often double-peaked, they are resolved spatially into two distinct offset \lya\ lines and corresponding continua in 2-D spectra. For HPS194 and HPS251, this may be an alternative explanation for the multiple-peaked \lya\ emission and why the expanding shell models poorly represent the data. However, our fiber-based \lya\ spectra are spatially unresolved on the sky, so we cannot investigate potential spatial offsets of the spectrally resolved \lya\ emission components for these galaxies with the current data. As seen in Figure \ref{fig:PointingThumbs}, the potential companion sources for HPS256 and the confirmed companion to HPS251 are located outside of the Mitchell Spectrograph's fiber, even when considering the RMS pointing uncertainty. Yet, we still observe the multiple-peaked \lya\ line profile morphology. This makes it more likely that the multiple-peaked and asymmetric nature of the \lya\ line for these galaxies is a consequence of radiative transfer effects in a non-static medium rather than being the result of the integrated \lya\ emission from multiple sources at similar redshift. 

Since all three galaxies have at least one nearby continuum source within a projected distance of $\lesssim33$ kpc and one of our galaxies has a \textit{confirmed} companion within $\sim8$ kpc, a possible link between galaxy-galaxy interactions and luminous \lya\ emission should not be ignored. This has recently been explored in the context of LABs by \citet{yajima2012} who post-process hydrodynamical simulations of gas-rich binary major mergers with a 3-D radiative transfer code. These authors find that such mergers produce copious \lya\ emission (with \lya\ luminosity $L_{\mathrm{Ly}\alpha} \sim 10^{43-44}$ erg s$^{-1}$) that is extended over 20-50 kpc at $z\sim3$ as a result of shocked gas and the starburst induced by the gravitational interaction. These properties are similar to typical $z\sim3$ LABs (e.g., \citealp{matsuda2006}). Due to the observational and selection methods utilized in the HPS \citep{adams2011}, the $L_{\mathrm{Ly}\alpha}$ observed for our three LAEs ($\sim 10^{43}$ erg s$^{-1}$; see Table \ref{table:GalData}) is of the same order as that predicted by the \citet{yajima2012} simulations. Additionally, the \textit{total} velocity widths of the \lya\ emission for HPS194, HPS256, and HPS251 are 667, 612, and 884 km s$^{-1}$, respectively \citep{adams2011}, which are each comparable to the median value of 780 km s$^{-1}$ for the LABs of \citet{matsuda2006}. However, since the \lya\ spectra we present in this work supply no spatial information, we cannot currently assess the extended nature of the three LAEs individually. The HPS data is also of limited use in this regard due to the coarse spatial resolution and limited depth of the survey (an upper limit of $7.5\arcsec$ FWHM, which corresponds to the spatial resolution limit of the HPS, can be placed on each LAE's size; \citealp{adams2011}). Future deep, spatially resolved \lya\ spectra will be useful in determining if our LAEs are significantly extended, which would provide further evidence for the interaction scenario based on the simulations by \citet{yajima2012}.

Recently, \citet{rhoads2013} used measurements of the SFR of bright LAEs (HPS194 and HPS256 were among the galaxies in their sample) along with measurements of their dynamical mass (which were assumed to be the upper limit of their gas mass) and sizes to compare such galaxies to existing star formation scaling relations. \citet{daddi2010} established that there are two distinct sequences of star formation in the Kennicutt-Schmidt $\Sigma_{\mathrm{SFR}}$ vs. gas mass density $\Sigma_{\mathrm{gas}}$ plane: a temporally extended sequence for "normal" star-forming disk galaxies, and a more rapid sequence for starburst galaxies that was determined from measurements of sub-millimeter and ultraluminous infrared galaxies (SMGs and ULIRGs, respectively). The latter scaling relation is offset above the former by 1.1 dex in the $\Sigma_{\mathrm{SFR}}$ vs. $\Sigma_{\mathrm{gas}}$ plane. This bimodality between the star formation sequences of starburst and normal star-forming galaxies is largely due to the use of a bimodal conversion factor $\alpha_{\mathrm{CO}} = M_{\mathrm{gas}} / L_{\mathrm{CO}}$ used to convert the observed carbon monoxide line luminosity $L_{\mathrm{CO}}$ to gas mass $M_{\mathrm{gas}}$ ($\alpha_{\mathrm{CO}} = 4.6$ for normal star-forming galaxies and 0.8 for starburst galaxies). The measurements by \citet{rhoads2013} suggest that LAEs lie above the scaling relation for normal star-forming galaxies and are consistent with the \citet{daddi2010} starburst sequence at the $>3\sigma$ level. Thus, LAEs appear to form stars more rapidly than a typical star-forming galaxy at a given $\Sigma_{\mathrm{gas}}$. Note that the bimodality of $\alpha_{\mathrm{CO}}$ has recently been challenged by the simulations of \citet{narayanan2012}. These authors show that while $\alpha_{\mathrm{CO}}$ is systematically different for varying local galaxy conditions (e.g., metallicity and surface density), it varies smoothly among them. The result is a continuous, unimodal star formation scaling relation where the highest $\Sigma_{\mathrm{SFR}}$ disk galaxies overlap with the inferred mergers (i.e., starburst galaxies). The measurements for LAEs by \citet{rhoads2013} show that LAEs occupy a region in the $\Sigma_{\mathrm{SFR}}$ vs. $\Sigma_{\mathrm{gas}}$ plane bounded by $0 \lesssim \log\Sigma_{\mathrm{SFR}} \lesssim 1.3$ and $2 \lesssim \log\Sigma_{\mathrm{gas}} \lesssim 3.2$. Data points within this bounded area are consistent with the scatter in the continuous star formation scaling relation of \citet{narayanan2012} and lie in a region of elevated SFR density occupied by both high-$z$ disk galaxies and inferred mergers (i.e., low-$z$ ULIRGs and high-$z$ SMGs). Thus, the star formation observed for luminous LAEs is consistent with (but not necessarily suggestive of) that expected from an interacting or merging system.

In this scenario, interactions could be responsible for dispersing gas and dust and allowing \lya\ photons to escape through low column density (i.e., low optical depth) ``windows'' in the overall neutral gas distribution. Combined with a low dust optical depth, the gravitationally induced burst of star formation could result in the large \ewlya\ we observe in addition to driving a strong large-scale outflow. \lya\ radiative transfer through the dispersed (i.e., lower \nh) outflowing gas could give rise to the asymmetric, multiple-peaked \lya\ line profiles with small \lya\ velocity offsets, as compared to the \lya\ spectra of LBGs that typically have large velocity offsets and smaller \ewlya. Interaction-induced inflows of gas (resulting in star formation or shocks) along the line of sight could also help to enhance the \lya\ flux blueward of the line center. Additionally, multiple blueward \lya\ peaks (such as that observed for HPS256 and HPS251) could also be produced from shocked and/or fluorescing gas that is infalling along different sight lines. The large blueshift ($\sim1000$ km s$^{-1}$) of these \lya\ peaks, however, likely indicates that such emitting material is not yet bound, unless radiative transfer effects couple favorably with the dynamics of the system to produce such large velocity offsets relative to the systemic velocity of the LAE's \hii\ regions. To verify an interaction-based scenario for luminous LAEs, a larger sample of galaxies with confirmed redshifts for nearby companion continuum sources and spatially resolved \lya\ spectra are needed. 

The irregular nature of these galaxies' continuum morphologies and the confirmed and potential interactions with nearby companions may suggest that significant deviations away from \lya\ point sources and spherical gas distributions and velocity fields are necessary for properly modeling \lya\ radiative transfer through neutral gas on the galactic and circumgalactic scale. This conjecture is especially intriguing given that the two galaxies that are most poorly represented by the spherical expanding shell model \lya\ spectra (i.e., HPS194 and HPS251) are the same two galaxies with the strongest evidence of an interaction.

\section{SUMMARY \& CONCLUSION}\label{sec:Conclusions}
In this paper, we have obtained follow-up optical spectra of three LAEs drawn from the HETDEX Pilot Survey \citep{adams2011} with sufficient spectral resolution to resolve the \lya\ emission line. With no preselection other than $F_{\mathrm{Ly}\alpha} > 10^{-16}$ \fluxcgs\ (which corresponds to $L_{\mathrm{Ly}\alpha} \gtrsim 10^{43}$ erg s$^{-1}$ at $z\sim2.4$), we find that all three galaxies at 120 km s$^{-1}$ FWHM spectral resolution display multiple \lya\ emission peaks. Using the NIR spectra of these galaxies' rest-frame optical emission lines from \citet{finkelstein2011} and Song et al. (2013, in preparation), we have determined the velocity structure of the \lya\ emission relative to the systemic redshift of each galaxy. Our main results are as follows:

1) The prominent double-peaks of the \lya\ emission line for each LAE straddles and is asymmetric about the velocity zero-point. The strongest emission peak is redshifted by 176 km s$^{-1}$ on average relative to the systemic velocity and its velocity offset and basic shape are strikingly similar among the three galaxies in our sample. We observe larger variations from galaxy to galaxy in the emission blueward of the systemic velocity, including two of the three galaxies that display two separate weak blueshifted emission peaks. The most blueshifted of these weak peaks for these two galaxies is offset from the systemic redshift by $\sim1000$ km s$^{-1}$. However, the velocity offset of the redward peak and the peak-to-peak velocity separation between the two most prominent peaks that straddle the velocity zero-point are $\sim2\times$ smaller for each LAE than the same measurements made on average for \lya\ emitting LBGs with similar \lya\ line profile morphologies. This is true even when taking into account the spectral resolution differences for each sample. 

2) We have compared our \lya\ spectra with the predicted line profiles of a grid of \lya\ radiative transfer models in the popular and relatively idealized spherical expanding shell geometry to model large-scale galactic outflows. In contrast to the findings of works such as \citet{verhamme2008} for single-peaked asymmetric \lya\ line profiles, we observe several key discrepancies between the best-fitting models and the data. Visually, the redshifted \lya\ emission component is acceptably reproduced by models with low column density of neutral gas ($N_{\mathrm{HI}} = 10^{18}$ cm$^{-2}$). However, the blueshifted emission component ubiquitously has too little flux as compared to the data and has an incorrect velocity offset relative to the systemic velocity for two of the three galaxies (HPS194 and HPS251). Additionally, \lya\ radiative transfer through an expanding shell cannot produce the highly blueshifted emission peak that is observed for HPS256 and HPS251.  

3) Based on the above analysis, we caution against equating an observed velocity offset of \lya\ directly with an outflow or inflow velocity, especially at low spectral resolutions where the \lya\ line profile is unresolved and the intrinsic asymmetry of the emission line can bias the velocity offset measurement. The measured velocity offset, especially for unresolved \lya\ spectra, is a complex function of many parameters describing the neutral gas, especially the column density \nh\ due to its large dynamic range and ability to significantly change the width and asymmetry of the emission components. Additionally, the \lya\ line profile shape and the resulting velocity offset can also be highly dependent on the viewing angle for more realistic, non-spherical gas geometries. 

4) For luminous LAEs with $L_{\mathrm{Ly}\alpha} \gtrsim 10^{43}$ erg s$^{-1}$, like those in the HPS LAE sample, galaxy-galaxy interactions may play a significant role in producing and aiding the escape of copious \lya\ photons and in shaping the emergent line profile by inducing star formation and clearing gas and dust. Such gravitational interactions may cause deviations away from spherical neutral gas geometries and velocity fields, such as those modeled in our expanding shell radiative transfer simulations. The effects of a non-spherical outflow in addition to simultaneous inflow of neutral gas, as well as smaller-scale \lya\ radiative transfer effects within the ISM that are not captured in the expanding shell model could all contribute to the discrepancies we observe between the \lya\ spectra of these galaxies and the predictions of the models. 

In forthcoming work, we will extend this study to obtain spectrally resolved \lya\ line profiles of a larger sample of $\sim30$ high \ewlya\ LAEs using multi-object techniques. Of particular importance in the forthcoming work will be the ability to obtain at least one dimension of spatial information. The combination of a larger sample and $>$1-D spectra will allow us to not only investigate the frequencies of various \lya\ line profile morphologies among this unique sample, but also investigate the spatial distribution and possible extended nature of the \lya\ emission. The latter may provide important constraints on the emission and escape mechanisms of \lya\ photons from these systems.\\

T.S.C. acknowledges the support of a National Science Foundation Graduate Research Fellowship during this work. This research is partially supported by the National Science Foundation under grant AST-0926815 and the Texas Norman Hackerman Advanced Research Program under grant 003658-0295-2007. The construction of the Mitchell Spectrograph was possible thanks to the generous support of the Cynthia \& George Mitchell Foundation. We thank McDonald Observatory and its staff for graciously supporting the observations. We would additionally like to thank the following individuals for useful scientific discussions: V. Bromm, M. Fumagalli, J. Greene, D. Jaffe, J. Jardel, E. Komatsu, K. Kulas, E. McLinden, C. Scarlata, C. Steidel, V. Tilvi, and S. Tuttle. Finally, we acknowledge the anonymous referee for useful comments that have improved this paper. \\



\end{document}